\documentclass[twocolumn]{aastex631}

\usepackage{amsmath}
\usepackage{hyperref}
\usepackage{booktabs}
\usepackage{longtable}
\usepackage{savesym}
\savesymbol{tablenum}
\usepackage{placeins}
\usepackage{siunitx}
\usepackage{natbib}
\usepackage[T1]{tipa}

\usepackage{threeparttable}
\restoresymbol{SIX}{tablenum}
\DeclareSIUnit{\parsec}{pc}
\DeclareSIUnit{\jansky}{Jy}
\DeclareSIUnit{\dmunit}{pc cm^{-3}}

\DeclareTextFontCommand{\textipa}{%
  \fontfamily{cmss}\tipaencoding
}

\newcommand{\kkoname}{k'ni\textipa{P}atn k'l$\left._\mathrm{\smile}\right.$stk'masqt}
\newcommand{\msbold}[1]{\textbf{#1}}

\newcommand{\mstar}{\ensuremath{M^*}}

\newcommand{\dm}{\ensuremath{\rm DM}}
\newcommand{\dmmw}{\ensuremath{{\rm DM}_\textrm{MW}}}
\newcommand{\dmmwism}{\ensuremath{{\rm DM}_\textrm{MW,ISM}}}
\newcommand{\dmmwcgm}{\ensuremath{{\rm DM}_\textrm{MW,CGM}}}
\newcommand{\dmcosmic}{\ensuremath{{\rm DM}_{\rm cosmic}}}
\newcommand{\avgdmcosmic}{\ensuremath{\langle {\rm DM}_{\rm cosmic}(z) \rangle}}
\newcommand{\dmh}{\ensuremath{{\rm DM}_\textrm{host}}}
\newcommand{\dmhm}{\ensuremath{{\rm DM}_\textrm{host}^\textrm{Macquart}}}
\newcommand{\dmhism}{\ensuremath{{\rm DM}_\textrm{h,ISM}}}
\newcommand{\dmhcgm}{\ensuremath{{\rm DM}_\textrm{h,CGM}}}
\newcommand{\dmhcbm}{\ensuremath{{\rm DM}_\textrm{CBM}}}

\newcommand{\dmtot}{\ensuremath{{\rm DM}_\textrm{FRB}}} 

\newcommand{\sigmacosmic}{\ensuremath{{\sigma}_\textrm{cosmic}}}

\received{\today}%
\shortauthors{Leung and the CHIME/FRB Collaboration et al.}
\graphicspath{{./}{./figures/}}

\begin{document}

\title{Stellar Mass-Dispersion Measure Correlations Constrain Baryonic Feedback in Fast Radio Burst Host Galaxies}

\shorttitle{Stellar Mass, Dispersion Measure, and Baryonic Feedback}
\author[0000-0002-4209-7408]{Calvin Leung}
  \affiliation{Department of Astronomy, University of California, Berkeley, CA 94720, United States}
  \affiliation{Miller Institute for Basic Research, University of California, Berkeley, CA 94720, United States}
\author[0000-0003-3801-1496]{Sunil Simha}
  \affiliation{Center for Interdisciplinary Exploration and Research in Astronomy, Northwestern University, 1800 Sherman Avenue, Evanston, IL 60201, USA }
  \affiliation{Department of Astronomy and Astrophysics, University of Chicago, William Eckhardt Research Center, 5640 S Ellis Ave, Chicago, IL 60637}
\author[0009-0000-7415-8239]{Isabel Medlock}
  \affiliation{Department of Astronomy, Yale University, New Haven, CT 06520, USA}
\author[0000-0002-6766-5942]{Daisuke Nagai}
  \affiliation{Department of Physics, Yale University, New Haven, CT 06520, USA}
  \affiliation{Department of Astronomy, Yale University, New Haven, CT 06520, USA}
\author[0000-0002-4279-6946]{Kiyoshi W.~Masui}
  \affiliation{MIT Kavli Institute for Astrophysics and Space Research, Massachusetts Institute of Technology, 77 Massachusetts Ave, Cambridge, MA 02139, USA}
  \affiliation{Department of Physics, Massachusetts Institute of Technology, 77 Massachusetts Ave, Cambridge, MA 02139, USA}
\author[0009-0007-5296-4046]{Lordrick A.~Kahinga}
  \affiliation{Department of Astronomy and Astrophysics, University of California, Santa Cruz, 1156 High Street, Santa Cruz, CA 95060, USA}
\author[0000-0003-2116-3573]{Adam E.~Lanman}
  \affiliation{MIT Kavli Institute for Astrophysics and Space Research, Massachusetts Institute of Technology, 77 Massachusetts Ave, Cambridge, MA 02139, USA}
  \affiliation{Department of Physics, Massachusetts Institute of Technology, 77 Massachusetts Ave, Cambridge, MA 02139, USA}
\author[0000-0002-3980-815X]{Shion Andrew}
  \affiliation{MIT Kavli Institute for Astrophysics and Space Research, Massachusetts Institute of Technology, 77 Massachusetts Ave, Cambridge, MA 02139, USA}
  \affiliation{Department of Physics, Massachusetts Institute of Technology, 77 Massachusetts Ave, Cambridge, MA 02139, USA}
\author[0000-0003-3772-2798]{Kevin Bandura}
  \affiliation{Lane Department of Computer Science and Electrical Engineering, 1220 Evansdale Drive, PO Box 6109, Morgantown, WV 26506, USA}
  \affiliation{Center for Gravitational Waves and Cosmology, West Virginia University, Chestnut Ridge Research Building, Morgantown, WV 26505, USA}
\author[0000-0002-8376-1563]{Alice P.~Curtin}
  \affiliation{Trottier Space Institute, McGill University, 3550 rue University, Montr\'eal, QC H3A 2A7, Canada}
  \affiliation{Department of Physics, McGill University, 3600 rue University, Montr\'eal, QC H3A 2T8, Canada}
\author[0000-0002-3382-9558]{B.~M.~Gaensler}
  \affiliation{Department of Astronomy and Astrophysics, University of California, Santa Cruz, 1156 High Street, Santa Cruz, CA 95060, USA}
  \affiliation{Dunlap Institute for Astronomy and Astrophysics, 50 St. George Street, University of Toronto, ON M5S 3H4, Canada}
  \affiliation{David A. Dunlap Department of Astronomy and Astrophysics, 50 St. George Street, University of Toronto, ON M5S 3H4, Canada}
\author[0000-0001-6128-3735]{Nina Gusinskaia}
  \affiliation{ASTRON, Netherlands Institute for Radio Astronomy, Oude Hoogeveensedijk 4, 7991 PD Dwingeloo, The Netherlands}
\author[0000-0003-3457-4670]{Ronniy C.~Joseph}
  \affiliation{Department of Physics, McGill University, 3600 rue University, Montr\'eal, QC H3A 2T8, Canada}
  \affiliation{Trottier Space Institute, McGill University, 3550 rue University, Montr\'eal, QC H3A 2A7, Canada}
\author[0000-0002-5857-4264]{Mattias Lazda}
  \affiliation{Dunlap Institute for Astronomy and Astrophysics, 50 St. George Street, University of Toronto, ON M5S 3H4, Canada}
  \affiliation{David A. Dunlap Department of Astronomy and Astrophysics, 50 St. George Street, University of Toronto, ON M5S 3H4, Canada}
\author[0000-0003-4584-8841]{Lluis Mas-Ribas}
  \affiliation{Department of Astronomy and Astrophysics, University of California, Santa Cruz, 1156 High Street, Santa Cruz, CA 95060, USA}
  \affiliation{University of California Observatories, 1156 High Street, Santa Cruz, CA 95060, USA}
\author[0000-0001-8845-1225]{Bradley W.~Meyers}
  \affiliation{International Centre for Radio Astronomy Research (ICRAR), Curtin University, Bentley WA 6102 Australia}
  \affiliation{Australian SKA Regional Centre (AusSRC), Curtin University, Bentley WA 6102 Australia}
\author[0000-0003-0510-0740]{Kenzie Nimmo}
  \affiliation{MIT Kavli Institute for Astrophysics and Space Research, Massachusetts Institute of Technology, 77 Massachusetts Ave, Cambridge, MA 02139, USA}
\author[0000-0002-8912-0732]{Aaron B.~Pearlman}
  \affiliation{Department of Physics, McGill University, 3600 rue University, Montr\'eal, QC H3A 2T8, Canada}
  \affiliation{Trottier Space Institute, McGill University, 3550 rue University, Montr\'eal, QC H3A 2A7, Canada}
  \affiliation{Banting Fellow}
  \affiliation{McGill Space Institute Fellow}
  \affiliation{FRQNT Postdoctoral Fellow}
\author[0000-0002-7738-6875]{J.~Xavier Prochaska}
  \affiliation{Department of Astronomy and Astrophysics, University of California, Santa Cruz, 1156 High Street, Santa Cruz, CA 95060, USA}
  \affiliation{Kavli Institute for the Physics and Mathematics of the Universe (Kavli IPMU), 5-1-5 Kashiwanoha, Kashiwa, 277-8583, Japan}
  \affiliation{Division of Science, National Astronomical Observatory of Japan, 2-21-1 Osawa, Mitaka, Tokyo 181-8588, Japan}
\author[0000-0002-4623-5329]{Mawson W.~Sammons}
  \affiliation{Department of Physics, McGill University, 3600 rue University, Montr\'eal, QC H3A 2T8, Canada}
  \affiliation{Trottier Space Institute, McGill University, 3550 rue University, Montr\'eal, QC H3A 2A7, Canada}
\author[0000-0002-6823-2073]{Kaitlyn Shin}
  \affiliation{MIT Kavli Institute for Astrophysics and Space Research, Massachusetts Institute of Technology, 77 Massachusetts Ave, Cambridge, MA 02139, USA}
  \affiliation{Department of Physics, Massachusetts Institute of Technology, 77 Massachusetts Ave, Cambridge, MA 02139, USA}
\author[0000-0002-2088-3125]{Kendrick Smith}
  \affiliation{Perimeter Institute of Theoretical Physics, 31 Caroline Street North, Waterloo, ON N2L 2Y5, Canada}
\author[0000-0002-1491-3738]{Haochen Wang}
  \affiliation{MIT Kavli Institute for Astrophysics and Space Research, Massachusetts Institute of Technology, 77 Massachusetts Ave, Cambridge, MA 02139, USA}
  \affiliation{Department of Physics, Massachusetts Institute of Technology, 77 Massachusetts Ave, Cambridge, MA 02139, USA}
\newcommand{\allacks}{
C. L. acknowledges support from the Miller Institute for Basic Research at UC Berkeley.
S.S. is supported by the joint Northwestern University and University of Chicago Brinson Fellowship.
I.M. and D.N. are supporeted by the NSF grant AST 2206055 and the Yale Center for Research Computing facilities and staff.
K.W.M. holds the Adam J. Burgasser Chair in Astrophysics and is supported by NSF grant 2018490.
A.P.C. is a Vanier Canada Graduate Scholar. 
K.N. is an MIT Kavli Fellow.
A.B.P. is a Banting Fellow, a McGill Space Institute~(MSI) Fellow, and a Fonds de Recherche du Quebec -- Nature et Technologies~(FRQNT) postdoctoral fellow.
M.W.S. acknowledges support from the Trottier Space Institute Fellowship program.
K.S. is supported by the NSF Graduate Research Fellowship Program.
}

\correspondingauthor{Calvin Leung}
\email{calvin\_leung@berkeley.edu}


\collaboration{99}{(CHIME/FRB Collaboration)}



\begin{abstract}
Low redshift fast radio bursts (FRBs) enable robust measurements of the host galaxy contribution to the dispersion measure (DM), offering valuable constraints on the circumgalactic medium (CGM) of FRB hosts.
We curate a sample of 20 nearby FRBs with low scattering timescales and face-on host galaxies with stellar masses ranging from $10^9 < M^* / M_\odot < 10^{11}$.
We fit the distribution of the host galaxy DM to a quadratic model as a function of stellar mass with a mass-independent scatter and find that the more massive the host, the lower its host DM.
We report that this relation has a negative slope of $m = -97 \pm 44$ pc/cm$^{3}$/dex in stellar mass. 
We compare this measurement to similar fits to three sub-grid models implemented in the CAMELS suite of simulations from Astrid, IllustrisTNG, and SIMBA and find that fine-tuning of the host ISM contribution as a function of stellar mass is required in order to reconcile the observational data with the predictions of the fiducial CAMELS-Astrid model.
More generally, models which attribute a positive correlation between stellar mass and host dispersion measure ($m > 0$) to the CGM are in tension with our measurement. 
We show that this conclusion is robust to a wide range of assumptions, such as the offset distribution of FRBs from their hosts and the statistics of the cosmic contribution to the DM budget along each sightline.
Our results indirectly imply a lower limit on the strength of baryonic feedback in the Local Universe $(z < 0.2)$ in isolated $\sim L^*$ halos, complementing results from weak lensing surveys and kSZ observations which target higher halo mass and redshift ranges.
\end{abstract}

\section{Introduction}\label{sec:intro}
Understanding the ionized gas in and around galaxies is a major missing puzzle piece for galaxy formation and cosmology. It plays a crucial role in the baryon cycle and is a major contaminant for nascent weak lensing surveys including the Vera Rubin Observatory~\citep{lsst2009science}, Euclid~\citep{euclid2025overview}, and the Nancy Grace Roman Space Telescope~\citep{spergel2015wide}. The constraining power of those surveys hinges on an accurate and precise understanding of the impact of baryonic feedback on the matter power spectrum~\citep[for example]{white2004baryons, zhan2004effect, rudd2008effects}.

The dispersion measure, or DM, of fast radio bursts~\citep{petroff2019fast,cordes2019fast,petroff2022fast} can precisely measure the column density of electrons between the observer and the FRB source~\citep{lorimer2004handbook}. 
This makes FRBs excellent probes of extragalactic ionized gas, sensitive to both the unbound baryons in the large scale structure~\citep{masui2015dispersion,madhavacheril2019cosmology} and the collapsed gas overdensities on the outskirts of galaxies and their dark matter halos~\citep{mcquinn2014locating,prochaska2019probing}.

Several classes of methodologies have emerged which use FRBs to probe extragalactic baryons. 
One pioneered by~\citet{macquart2020census} aims to model the relationship between the redshift of the FRB host and the total extragalactic DM. The tight relationship between dispersion measure and redshift is known as the Macquart relation \avgdmcosmic\ and now extends to $z > 1$~\citep{connor2024gas}. It implies that other contributions, such as those from the host galaxy or individual groups and clusters, are subdominant to~\avgdmcosmic\ beyond $z \gtrsim 0.5$.

Other analyses aim to isolate particular contributions to the FRB DM, which is the sum of contributions from the Milky Way, overdensities and underdensities in the large scale structure, intervening halos, the host galaxy of the FRB, and the circumburst environment.

For instance, angular cross-correlations with galaxy surveys ~\citep{shirasaki2017large, madhavacheril2019cosmology,rafiei_ravandi2021chime,shirasaki2022probing,hsu2025decoding,wang2025measurement} and weak lensing~\citep{nicola2022breaking,reischke2023calibrating} are sensitive to baryons in the large-scale structure. Spectroscopic mapping of FRB foregrounds~\citep{lee2022constraining,lee2023frb,simha2020disentangling,simha2021estimating,khrykin2024flimflam,huang2025frb} can infer the contributions from individual galactic halos~\citep{prochaska2019probing}, groups, and clusters~\citep{connor2023deep} along FRB sightlines to the DM budget to separate the bound and the diffuse baryons along the line of sight. Edge-on FRB hosts probe the interstellar medium (ISM) of the host galaxy~\citep{cassanelli2024fast}, whereas all-sky FRB surveys statistically place upper limits on the Milky Way's halo~\citep{cook2023frb}.

Schematically, the DM budget can be modeled as 

\begin{equation}
\dmtot = \dmmw + \avgdmcosmic + \delta\dm_{LSS}(z) + \dfrac{\dmh}{1 + z}.
\label{eq:dmh_definition}
\end{equation}
where $\dmmw=\dmmwism + \dmmwcgm$ consists of contributions from the interstellar medium (ISM) and circumgalactic medium (CGM) of the Milky Way. \avgdmcosmic~is the average cosmic contribution to the DM budget out to redshift $z$, specified by the Macquart relation. Underdensities and overdensities due to large-scale structure along the line of sight, groups and clusters, and intervening galaxies contribute to $\delta\dm_{LSS}$. 

The last term is considered to be the ``host contribution'' to the DM budget. \dmh~can be further decomposed in the rest frame of the host:
\begin{equation}
    \dmh = \dmhcgm + \dmhism + \dmhcbm.
    \label{eq:dmh_contributions}
\end{equation}
The three terms here correspond to the CGM and the ISM of the host galaxy and contributions from the stellar ``circum-burst medium'' (CBM) of the FRB~\citep[see e.g.][]{tendulkar2017host, niu2021repeating,ocker2022large,caleb2023subarc}. 

The ``Macquart'' estimate of the host DM is given by solving Eq.~\ref{eq:dmh_definition} for \dmh, and using models for the Milky Way ISM~\citep{cordes2002ne2001} and CGM~\citep{yamasaki2020galactic} and the Macquart relation to estimate the Galactic and cosmic contributions respectively. 
\begin{equation}
\dfrac{\dmhm}{1+z} \approx \dmtot - \dm_\mathrm{NE2001} - \dm_\mathrm{YT20} - \avgdmcosmic.
\end{equation}

Because of the stochastic large-scale structure contribution and uncertainties in Galactic electron models, individual estimates of \dmhm\ are noisy: they are either positive or negative. However, measurements of $\langle \dmhm \rangle$ tend to scatter around a positive mean~\citep{shannon2018dispersion,macquart2020census,james2022fast,shin2023inferring,bernalescortes2025empirical}.



In this paper, we use statistics of \dmhm\ to constrain the average circumgalactic contribution of the host galaxy and its scatter as a function of stellar mass \mstar. 
In \S~\ref{sec:sample}, we consider a sample of \dmhm\ measurements selected to minimize contributions to \dmhm\ other than $\dmhcgm$, by using bursts away from the Galactic plane with low scattering timescales and whose hosts are nearby, do not belong to a galaxy cluster, and are viewed in a face-on geometry. Then we characterize the $M^*-\dmh$ relation for our sample of FRB host galaxies, in analogy to ``cluster scaling relations'' which correlate baryonic observables in more massive halos~\citep{battaglia2012on,giodini2013}. 

In \S~\ref{sec:camels}, we compare our measurements to predictions for \dmhcgm, the circumgalactic contribution to the host DM, from three sub-grid models implemented in CAMELS: a suite of small-volume (25 Mpc/h)$^3$ hydrodynamic simulations implementing several distinct sub-grid ``prescriptions'' found in the Astrid, IllustrisTNG, and SIMBA simulations~\citep{camels_presentation, camels_data_release,  camels_data_release2}. 
These prescriptions aim to accurately capture the large-scale impact of the unresolved small-scale physics that regulate galaxy formation and evolution. 
These include prescriptions for star formation, gas cooling, and feedback from stars and active galactic nuclei (AGN). 
One major goal of the CAMELS project is to produce theory predictions for gas probes which can independently constrain the suppression of the matter power spectrum induced by baryonic feedback relative to a dark-matter only simulation, as measured in current and next-generation weak lensing surveys such as the Dark Energy Survey~\citep{chen2023constraining,aric_o2023des} and LSST/Rubin. In \S~\ref{sec:analysis}, we compare our observations to simulations, and find that in order to reconcile the Astrid CGM prediction with the data, a negative correlation between the host ISM contribution and host stellar mass is needed. On the basis of this being potentially unphysical, we consider in \S~\ref{sec:discussion} the implications of the fiducial CAMELS-Astrid model being disfavored in the context of baryonic feedback in the Local Universe.

\section{Constructing the observed $M^*-\dmh$ relation}\label{sec:sample}
We aggregate all known FRB host galaxies to characterize the $M^*-\dmh$ relation for isolated, star forming galaxies.
These include sources from~\citet{macquart2020census,bhandari2020host,lanman2022sudden,leewadell2023host,michilli2023subarcminute,ravi2023deep,gordon2023demographics,bhardwaj2024host,law2024deep,sharma2024preferential,connor2024gas,cassanelli2024fast,shannon2024commensal,chime2025catalog}.
We impose a redshift cutoff of $z < 0.2$ to balance a trade-off:
on one hand, a cutoff that is more distant adds statistical constraining power by including more sources.
On the other hand, a less distant cutoff adds several advantages.
It makes our measurement of $\dmhcgm$ robust against redshift evolution and the cosmic contribution, whose mean \avgdmcosmic\ and scatter \sigmacosmic\ will add extra noise that also depends on feedback and increases with increasing redshift~\citep{jaroszynski2019fast,batten2021cosmic, walker2024dispersion}. 
It also reduces the fraction of putative host galaxies that are fainter than the depth of optical observations, reducing optical selection effects~\citep{seebeck2021effects,marnoch2023unseen,jahnsschindler2023how,hewitt2024repeating}.

\begin{figure*}
    \centering
    \includegraphics[width=\linewidth]{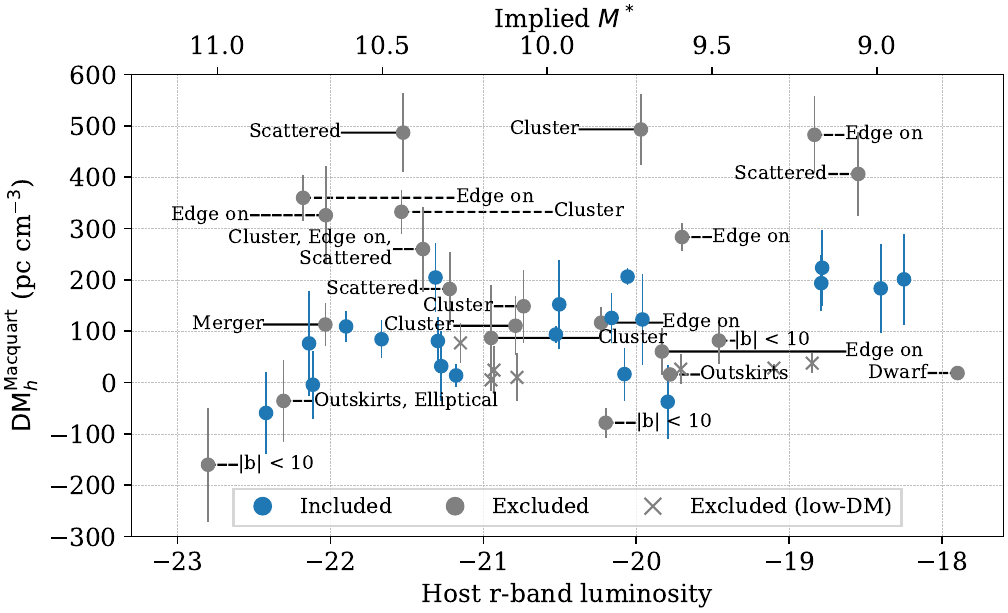}
    \caption{Measurements of $\dmhm$, as a function of $r$-band host galaxy luminosity for FRBs at $z < 0.2$ in the literature. The stellar mass implied by the~\citet{mahajan2018galaxy} conversion from r-band light to $M^*$~\citep{mahajan2018galaxy} is shown on the top y-axis to provide a rough sense of scale. For each source we plot $1\sigma$ error bars according to Eq.~\ref{eq:sigma_i_def}. Labels identify systems cut from our sample according to one or more criteria in \S~\ref{sec:sample}. We find that high \dmhm~outliers can often be explained astrophysically by high scattering timescales, cluster membership of the host galaxy, or edge-on host galaxies. Low \dmhm~outliers are often close to the Galactic plane. ``Low-DM'' FRBs, which have downwardly-biased DM values, are shown with gray X's.}
    \label{fig:host_dm}
\end{figure*}

For each FRB, we calculate $\dmhm$ according to Eq.~\ref{eq:dmh_definition}, using the NE2001~\citep{cordes2002ne2001} model to estimate $\dmmwism$ and the~\citet{yamasaki2020galactic} model to estimate $\dmmwcgm$, and the open-source\footnote{\texttt{average\_DM} function in \texttt{https://github.com/FRBs}; version 0.1} implementation of $\langle \dmcosmic(z) \rangle$.
All quantities in this work are calculated using the \textit{Planck} cosmology for a flat $\Lambda$CDM universe of $H_0 = 67.7~\mathrm{km~s}^{-1}~\mathrm{Mpc}^{-1}$, $\Omega_m=0.31$ and $\Omega_\Lambda = 0.69$ \citep{collaboration2020planck}. 
Where available, we use published stellar mass measurements derived from spectral energy distribution (SED) fitting. For fits using a Kroupa initial mass function (IMF)~\citep{gordon2023demographics,sharma2024preferential,connor2024gas} we scale stellar masses by 0.03 dex~\citep{madau2014cosmic} to standardize them to a~\citet{chabrier2003galactic} IMF.
For the ten sources that lack $M^*$ measurements from SED fitting,
we fit publicly available photometry from GALEX, DECaLS, PanSTARRS, and WISE using CIGALE~\citep{boquien2019cigale} assuming a Chabrier IMF and a delayed-$\tau$ star formation history. 
These fits converged for eight of the ten targets. 
The remaining two galaxies have stellar mass measurements from the NASA Extragalactic Database-Local Volume Survey~\citep[NED-LVS;][]{cook2023completeness} derived from a mass-to-light ratio assuming a Salpeter IMF.
We have validated that the stellar masses derived using CIGALE are consistent with the linear relation between $M^*$ and r-band light ($M_r$) characterized by~\citet{mahajan2018galaxy}. They differ by at most 0.5 dex over the range $10^8 < M^*/M_\odot < 10^{11}$, which is consistent with the scatter on that relation ($\pm 0.45$ dex at 1$\sigma$). The NED-LVS mass estimates are also consistant with the CIGALE estimates within fitting uncertainties for the two sources for which both data are available once the correction factor between a Salpeter and Chabrier IMF is applied.
\dmh~for all FRB sightlines at $z < 0.2$ is plotted as a function of $r$-band luminosity in Fig.~\ref{fig:host_dm}. 

From the data in Fig.~\ref{fig:host_dm}, we apply cuts to minimize contributions besides \dmhcgm to Eq.~\ref{eq:dmh_contributions}. The data points which pass these cuts are shown in blue in Fig.~\ref{fig:host_dm}, while data points that are cut are shown in gray. We cut the following systems:
\begin{enumerate}
    \item Host galaxies that are embedded in massive galaxy clusters including FRB 20220914A and FRB 20220509G~\citep{connor2023deep}, FRBs 20231206A, 20230203A, and 20231204A~\citep{chime2025catalog} (see also Lanman et al. in prep). The presence of a galaxy cluster in general may add several hundred pc cm$^{-3}$ to \dmhm\ and can explain some of the highest $\dmhm$ data points in Fig.~\ref{fig:host_dm}. We cross-match our sightlines with the cluster catalogs described in~\citet{tempel2018bayesian},~\citet{wen2018catalogue}, and~\citet{xu2022catalog} to check that their hosts are not within one cluster radius ($R_\mathrm{200,crit}$ or $R_\mathrm{500,crit}$ as reported by each catalog) of any cluster candidate within $\Delta z = 0.01$ of the host redshift. 
    The catalog footprints of~\citet{tempel2018bayesian} and~\citet{bahk2024galaxy} cover all of our sightlines;~\citet{xu2022catalog} covers 16/20 sightlines; and~\citet{wen2018catalogue} covers 13/20 sightlines.
    For the remaining sightlines we are unable to assess cluster membership without explicitly using DM-related information. However, the absence of large outlier \dmhm\ values in Fig.~\ref{fig:host_dm} and the rarity of $M_h \gtrsim 10^{12} M_\odot$ halos~\citep{mcquinn2014locating} suggest that hosts in clusters do not dominate the remaining \msbold{$\sim 50\%$} of sightlines. 
    
    For galaxy groups, we cross-match our hosts with~\citet{tully2015galaxy}, but are limited by its stringent redshift horizon of $<10000$ km/s. However, we exclude the galaxy-merger host of FRB 20190303A (noting that its \dmhm\ is not anomalously large).
    \item Host galaxies more inclined than $i = 70^\circ$. The host galaxy ISM is highly uncertain, so we minimize its contribution by excluding systems that are close to edge-on. Since there is significant uncertainty in inclination angle measurements, we only use it as a binary classifier to clearly define our sample. Some systems have published inclination measurements, including FRB 20210603A \citep[$i \approx 83^\circ$; ][]{cassanelli2024fast}, FRB 20220207C \citep[$i \approx 74-80^\circ$;][]{law2024deep,bhardwaj2024selection}. FRB 20240201A ($i \approx \cos^{-1}(b/a) \approx 80^\circ$) and FRB 20240310A ($i \approx 85^\circ$) are presented in~\citet{shannon2024commensal,gordon2025mapping}, and for these inclination estimates we use either inclination angles or galaxy minor and minor axes from~\citet{glowacki2025investigation} via $\cos(i) \approx b/a$. 
    For FRB 20231120A ($i \sim 81^\circ$)~\citep{sharma2024preferential}, FRB 20230203A ($i \sim 77-81^\circ$) and FRB 20231005A ($i \sim 81^\circ$), which lack published estimates, we apply the InclinationZoo estimator~\citep{kourkchi2020cosmicflows} to publicly-available SDSS and Pan-STARRS imaging, verifying that that measurements are consistent between both the shallow SDSS and deeper Pan-STARRS images. Where quoted, the range presented reflects the discrepancy between the central value of the SDSS and the Pan-STARRS inclination estimates). Inclination estimates of these systems fall decisively outside of the boundary for our sample, making inclination measurement uncertainties unlikely to significantly impact our results. 
    \item Host galaxies whose FRBs lie on the outskirts of their stellar disks. FRB 20200120E~\citep{bhardwaj2021nearby, kirsten2022repeating} and 20240209A~\citep{shah2024repeating} are located $\sim 5 R_\mathrm{eff}$ from their host galaxy centers. They may systematically underestimate the average host galaxy CGM.
    \item Dwarf and elliptical FRB hosts. While galaxy classification can be ill-defined, we exclude dwarfs and elliptical from our sample using a stellar mass cut ($10^9 < M^*/M_\odot < 10^{11}$). This cut excludes FRB 20240209A \citep{eftekhari2024massive}, the first and only unambiguously elliptical host to date, and FRB 20121102A~\citep{spitler2016repeating,tendulkar2017host}, whose host is a low-mass dwarf galaxy. The other two of the three published dwarf hosts~\citep[FRB 20190520B \& 20210117A][]{bhandari2023nonrepeating,hewitt2024repeating} are beyond $z > 0.2$, so they are already cut by our redshift horizon. The gas properties of dwarf and elliptical galaxies likely differ significantly from FRB hosts, which are typically (but not always) star-forming~\citep{bhandari2020host,glowacki2023wallaby,sharma2023deep}. We note that inclusion of the high \dmhm\ values of these three systems would only strengthen the measured anti-correlation between stellar mass and \dmhm (see \S~\ref{sec:analysis}).
    \item Sightlines at a Galactic latitude $|b_\mathrm{Gal}| < 10^\circ$. The Milky Way ISM adds noise due to the 20\% uncertainty on the DM$_\mathrm{NE2001}$ estimate of the DM. This cut excludes the hosts of FRB 20180916B~\citep{collaboration2020periodic,marcote2020repeating}, FRB 20220319A~\citep{ravi2023deep}, and FRB 20210405I~\citep{driessen2024frb}. 
    \item So-called ``low-DM'' FRB sightlines. For surveys with large ($\gtrsim 1'$) localization regions, the low total DM can be used to make a host galaxy association, but only by imposing a DM-inferred distance cutoff. However, these associations bias the sample because of the explicit cutoff on DM$_{tot}$, which underestimates $\dmhm$. There are nine FRB hosts identified this way described in~\citet{bhardwaj2021local,michilli2023subarcminute,bhardwaj2024host,Ibik2024host}: They include FRBs 20180814A, 20181030A, 20191220A, 20181223C, 20190418A, 20190110C, 20190425A, 20200223B, as well as 20190303A (note its repeat burst FRB 20231204A). 
    \item Hosts whose FRB scattering timescales exceed 5 ms at 600 MHz (extrapolated using a $\nu^{-4}$ scaling if necessary). The presence of large circum-burst contributions to the FRB DM may be indicated by scattering timescales~\citep{cordes2016radio, ocker2022large,caleb2023subarc} and in some cases the rotation measures as well~\citep{michilli2018extreme,mannings2023}, but we do not apply any cut on the latter. This cut excludes FRB 20190608B~\citep{chittidi2021dissecting}, FRB 20210410D~\citep{caleb2023subarc}, FRB 20230203A and FRB 20230222A~\citep{chime2025catalog} from our sample. 
\end{enumerate}

Applying these cuts leaves 20 FRB sightlines. Their \dmhm measurements and stellar masses are shown in Table~\ref{tab:sample}.

\begin{table}
\centering
\caption{A list of the FRBs used in this work according to the criteria described in Sec. 2. Letters denote sources of the stellar mass measurements used in this work: In order of priority we report $^{a}$ Published stellar masses, $^{b}$ New SED fits (uncertainties of $\leq 0.16$ dex). $^{c}$ NED-LVS stellar masses, all standardized to a Chabrier IMF (uncertainties of $0.3$ dex).}
\label{tab:sample}
\begin{tabular}{lrrr}
\toprule
TNS Name & $z_{spec}$ & $\log_{10}(M^*/M_\odot)$ & $\dmhm$ \\
\midrule
FRB 20200430A & 0.1608 & 9.51$^{a}$ & 201 \\
FRB 20210807D & 0.1293 & 10.94$^{a}$ & -59 \\
FRB 20211127I & 0.0469 & 9.45$^{a}$ & 109 \\
FRB 20211212A & 0.0715 & 10.25$^{a}$ & 81 \\
FRB 20220725A & 0.1926 & 10.71$^{c}$ & 76 \\
FRB 20220912A & 0.0771 & 10.00$^{a}$ & 194 \\
FRB 20220920A & 0.1582 & 9.81$^{a}$ & 123 \\
FRB 20230222B & 0.1100 & 10.19$^{c}$ & 32 \\
FRB 20230526A & 0.1570 & 9.31$^{b}$ & 183 \\
FRB 20230628A & 0.1265 & 9.26$^{a}$ & 224 \\
FRB 20230926A & 0.0553 & 10.49$^{b}$ & 84 \\
FRB 20231011A & 0.0783 & 9.59$^{b}$ & 16 \\
FRB 20231123A & 0.0729 & 9.42$^{b}$ & 126 \\
FRB 20231128A & 0.1079 & 9.47$^{b}$ & 205 \\
FRB 20231201A & 0.1190 & 9.47$^{b}$ & -38 \\
FRB 20231223C & 0.1059 & 10.40$^{c}$ & -4 \\
FRB 20231226A & 0.1569 & 9.59$^{b}$ & 152 \\
FRB 20231229A & 0.0190 & 9.87$^{bc}$ & 94 \\
FRB 20231230A & 0.0298 & 10.04$^{bc}$ & 14 \\
FRB 20240210A & 0.0237 & 9.67$^{b}$ & 206 \\
\bottomrule
\end{tabular}
\end{table}

The uncertainty of each individual \dmhm\ estimate is modeled as
\begin{equation}
    \sigma_{i}^2 = \sigmacosmic(z_i)^2 + (0.2 \dm_\mathrm{i,NE2001})^2,
    \label{eq:sigma_i_def}
\end{equation}

where we use a prescription for \sigmacosmic(z) measured from snapshots taken from large-volume hydrodynamic simulations that more accurately model large-scale structure not captured in the small CAMELS volumes.
The choice of \sigmacosmic(z) varies as a function of feedback model.
We linearly interpolate the IllustrisTNG values provided in Table 4 of~\citet{walker2024dispersion} in our fiducial analysis, but we test the impact of varying \sigmacosmic\ on our results, confirming that our measurements are insensitive to shifts in \sigmacosmic(z).
Furthermore, while some studies have suggested a non-Gaussian distribution of various components of the DM budget based on hydrodynamic simulations and analytical models~\citep{mcquinn2014locating,medlock2024probing,reischke2024analytical,sharma2025hydrodynamical}, the unknown contributions of the host ISM and Milky Way ISM complicate the picture.
We use Gaussian data uncertainties, which in the limit of multiple independent contributions to the DM budget, is most agnostic of specific model prescriptions for e.g. the ionized CGM and ISM.

\section{Predictions from CAMELS}\label{sec:camels}
We now aim to compare the data to theory predictions, built from a representative sample of sightlines from CAMELS.
We make our predictions by extending the ray tracing procedure of~\citet{medlock2024probing}.
We use both halo catalogs and simulation snapshots from the \texttt{1P} set of simulations in CAMELS. For simplicity, in this work we only consider the fiducial feedback prescription $(A_{SN1} = A_{SN2} = A_{AGN1} = A_{AGN2} = 1)$ at $z = 0.1$, the snapshot closest to the median redshift of our sample.
The halo catalogs are used to assign halo masses, stellar masses, and stellar radii to each halo in the snapshot.
Next, to minimize contamination from the host ISM, which is poorly resolved at the resolution of CAMELS, gas particles within $0.1 R_\mathrm{200,crit}$ of each halo in the halo catalog are removed from each snapshot.
The snapshots are then processed using \texttt{yt}~\citep{turk2011yt} to interpolate the electron density field onto a regular grid. 
The electron density is then integrated along one of the grid axes over the 25 Mpc/h extent of the snapshot, and divided by two to account for the fact that observed FRB sightlines skewer halfway through their host halos on average.
This integral over a 25 Mpc/h sightline within a snapshot defines \dmhcgm\ for this paper. 
We argue that for comparisons between simulated and observed FRB sightlines, this definition is better-suited than definitions which fix a boundary between the CGM and IGM as a function of halo mass, which is hard to accurately infer. 
One potential drawback is that our definition incurs a bias due to the mean gas density within the $25$ Mpc/h snapshot, but this on the order of $\lesssim 5$ pc cm$^{-3}$~\citep{medlock2024probing}. This bias can be subtracted, and with a $1/\sqrt{N}$ scaling of the uncertainties used in this work (Eq.~\ref{eq:sigma_i_def}), it is subdominant for fewer than $ N \lesssim 200$ data points.
After removing sightlines intercepting more than one halo within its $R_\mathrm{200,crit}$, like in criteria 2 in \S~\ref{sec:sample}, each sightline can be unambiguously associated with the index of the nearest halo in the halo catalog.
Then, within a range of normalized impact parameters $\rho_{\rm min} < \rho < \rho_{\rm max}$, where $\rho = r_\perp/R_\mathrm{200,crit}$, one sightline per halo is then selected at random.
This corrects for a preferential sampling of massive halos proportional to their larger two-dimensional cross-section within the snapshot.

\begin{figure*}
    \centering
    \includegraphics[width=1.0\linewidth]{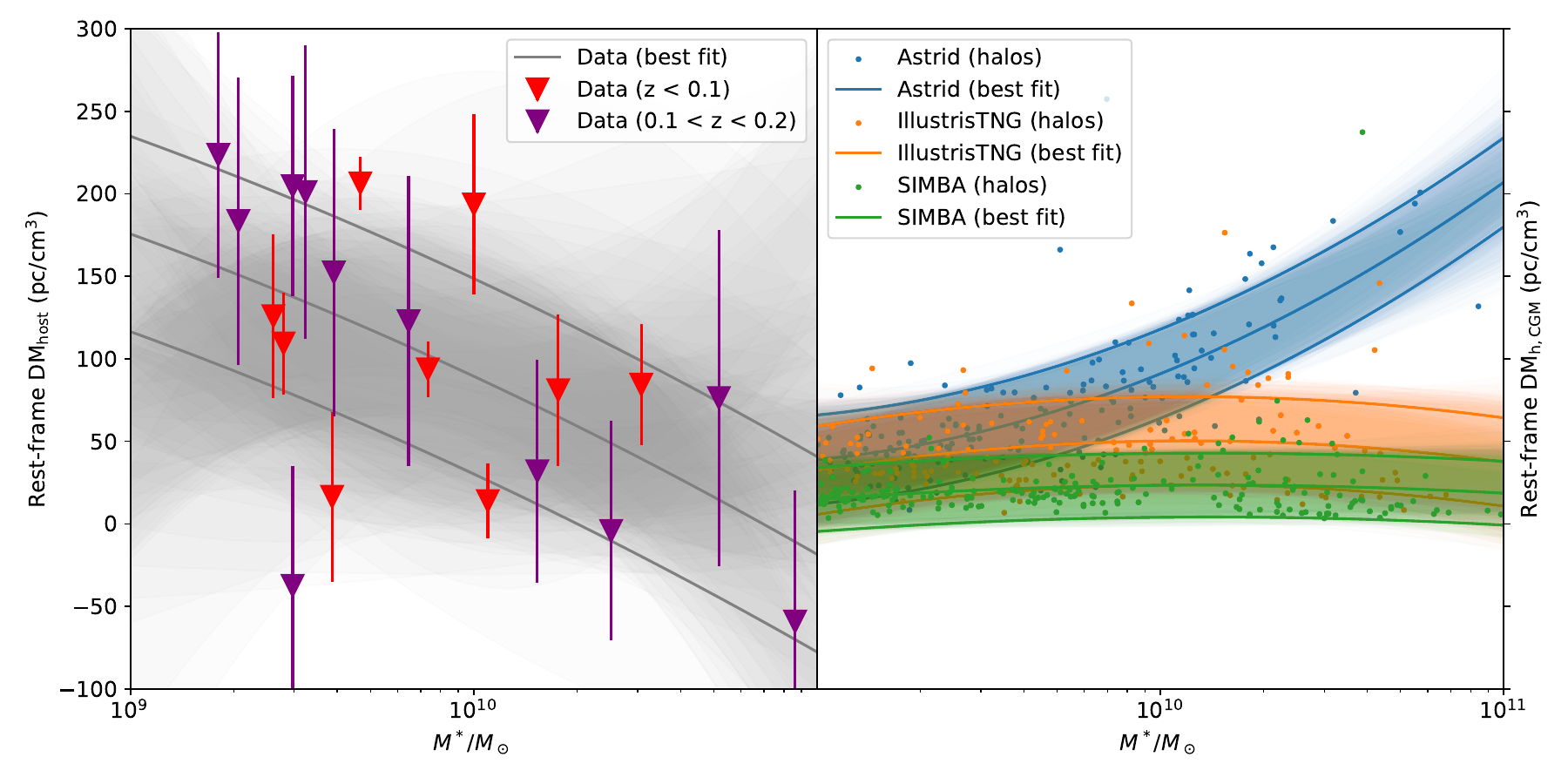}
    \caption{Left: The $M^*-\dmh$ relation for the FRB host galaxies in Tab.~\ref{tab:sample} (red and purple downward triangles), to which we fit the model in Eq.~\ref{eq:obs_model} (gray).
    Right: The $M^*-\dmhcgm$ prediction for halos in the CAMELS-ASTRID, CAMELS-IllustrisTNG, and CAMELS-SIMBA simulations (blue, orange, and green points). 
    For the fits to observational data and simulations, the model prediction best fitting the data is visualized as three lines, to show the relation as its scatter $\pm 1\sigma$.
    In addition, model evaluations randomly sampled from the posterior probability distributions are plotted translucently in those same colors.}
    \label{fig:camels}
\end{figure*}

In Fig.~\ref{fig:camels} we plot the \dmhcgm\ values for our simulated sightlines within a fiducial range of impact parameters ($\rho_{\rm min} = 0.0 < \rho < 0.05 = \rho_{\rm max}$) as a function of stellar mass and feedback model. These predictions are slightly sensitive to the host offset distribution of FRBs: FRB sightlines highly offset from their gas halos may exclude a small portion of \dmhcgm\ from their sightlines~\citep[see Fig. 5 of][]{medlock2024probing}. Though highly-offset FRB sources at impact parameters of $\sim 5 R_\mathrm{eff}$ have been associated with their host galaxies~\citep{kirsten2022repeating,shah2024repeating,eftekhari2024massive}, these sources have been cut from our observational sample (criteria 3 in \S~\ref{sec:sample}); most FRBs have offsets of less than $\approx 2 R_\mathrm{eff}$~\citep[see Fig. 6 in][]{shannon2024commensal}. Given that the half-light radius $R_\mathrm{eff}$ is $\sim 0.015 R_\mathrm{200,crit}$\citep{kravtsov2013size}, we may conclude that $\rho \lesssim 0.03 R_\mathrm{200,crit}$ for FRB sightlines in Tab.~\ref{tab:sample}, even though the second component of the host offsets of some sources is not well-constrained by their highly elongated localization ellipses~\citep{chime2025catalog}.

\section{Results}\label{sec:analysis}
We characterize the $M^*-\dmh$ relation by fitting both the observed data and the simulation data to a quadratic in $x = \log_{10}(M^*/(10^{10} M_\odot))$. 
We choose a quadratic functional form for the mean of the relation because a linear model was found to be insufficient to fit the simulation predictions (see Table~\ref{tab:params}, which shows evidence for non-zero curvature in most simulation suites).
We parameterize the mean of the relation using an intercept parameter $b$, a slope parameter $m$, and a quadratic term $a$.
To allow for scatter around this mean, we allow the distribution of \dmh\ to be scattered around its mean value. We parameterize the statistical distribution of \dmh\ as a Gaussian distribution. The mean of the distribution is a quadratic in stellar mass and the variance of the distribution is a $\sigma^2$, taken as a constant over stellar mass for simplicity. 

In the case of the simulation data, our model predicts only the mean and scatter of the CGM contribution to the DM:
\begin{equation}
    \dmhcgm(M^*) \sim \mathcal{N}(a x^2 + m x + b , \sigma^2)\label{eq:sim_model}
\end{equation}

In the case of the observational data, which contain all four components of Eq.~\ref{eq:dmh_contributions}, we fit
\begin{equation}
    \dmhm(M^*) \sim \mathcal{N}(a x^2 + m x + b, \sigma^2 + \sigma_i^2)\label{eq:obs_model}.
\end{equation}

When fitting the observational data, $\sigma_i$ accounts for $\sigma_\mathrm{cosmic}$ and the uncertainty in the Milky Way contribution to the DM, as defined in Eq.~\ref{eq:sigma_i_def}. Therefore, $\sigma^2$ in the data fit characterizes not only the scatter in the host CGM contribution, but also the scatter in the host ISM and circum-burst contributions as well.

For both observed and simulated data, we fit the model using maximum likelihood estimation with a Gaussian likelihood. Abbreviating the observational measurements of $\dm^\mathrm{Macquart}_\mathrm{host,i}$ as $y_\mathrm{i}$ and the model prediction from Eq.~\ref{eq:obs_model} as $\hat{y}_\mathrm{i}$, the likelihood is

\begin{equation}
    -\log \mathcal{L} = \dfrac{1}{2} \sum_{i} \dfrac{(y_\mathrm{i} - \hat{y}_{i})^2}{(\sigma^2 + \sigma_i^2)} - \log(2\pi(\sigma^2 + \sigma_i^2)).
\end{equation}
We use the same likelihood with $\sigma_i = 0$ for the simulation fits with $\dm_\mathrm{h,CGM,i}$ as $y_\mathrm{data,i}$. 

The Markov Chain Monte Carlo (MCMC) sampler \texttt{emcee}~\citep{foreman_mackey2013emcee} is used to sample the posterior distribution of the model parameters. We adopt uninformative priors on the parameters, with $\pi(a) \sim \mathcal{U}(-1000,1000)$ pc/cm$^{3}$/dex$^2$; $\pi(m) \sim \mathcal{U}(-300,300)$ pc/cm$^{3}$/dex;  $\pi(b) \sim \mathcal{U}(-150,250)$ pc/cm$^{3}$, and $\pi(\sigma) = 1/\sigma$ over the range $[1,200]$ pc/cm$^{3}$.

\begin{figure*}
    \centering
    \includegraphics[width=\textwidth]{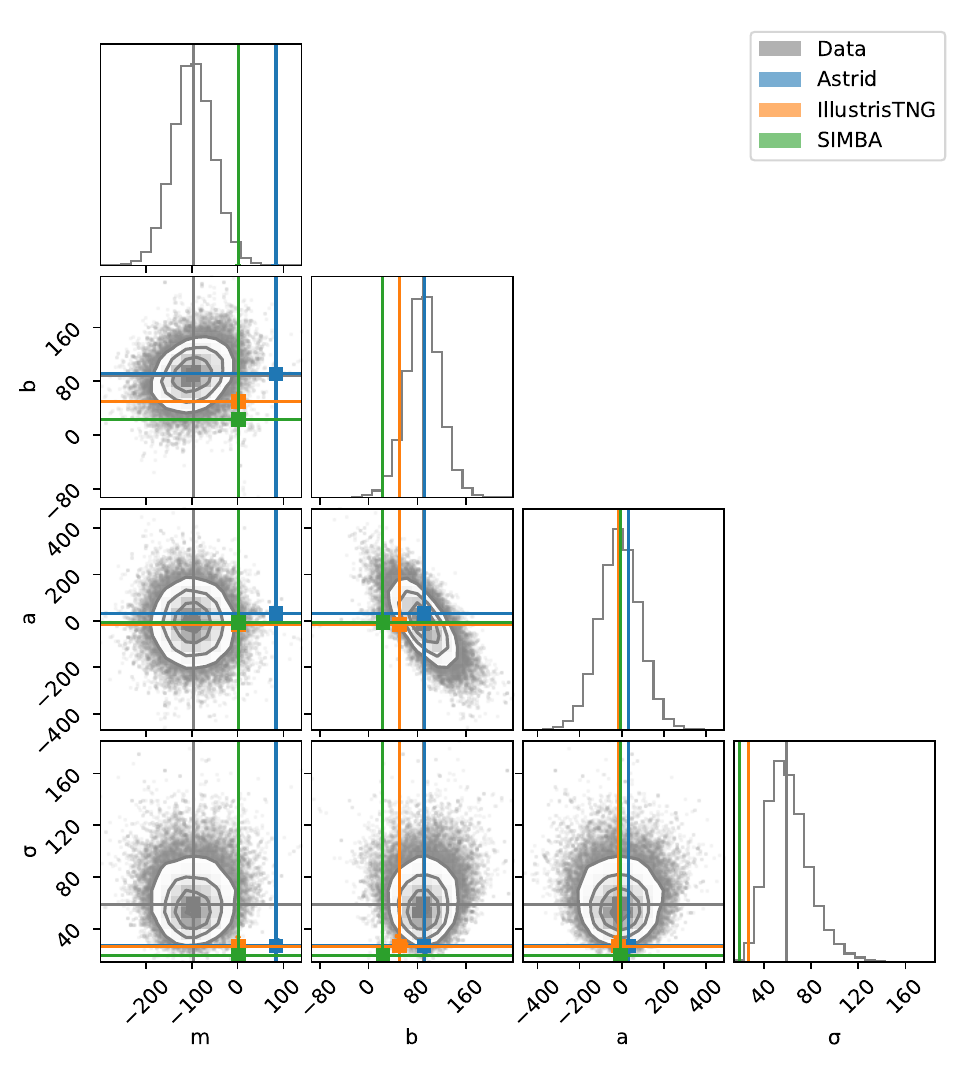}
    \caption{Posterior distributions of the model parameters in our fiducial analysis using all 20 bursts. We fit the $M^*-\dmh$ relation for a slope $(m)$, an intercept $(b)$ referenced to $M^* = 10^{10} M_\odot$, a quadratic curvature parameter $(a)$, as well as the scatter on the relation $(\sigma)$ (see Eq.~\ref{eq:sim_model},~\ref{eq:obs_model}). We fit both the observed sightlines (gray) and the sightlines simulated through CAMELS snapshots (blue, orange, and green). The best-fit parameter estimates are shown as solid lines, and are quoted in Tab.~\ref{tab:params}. Comparing the data to the Astrid fits, we find that the slope parameter $m$ is significantly discrepant between observations and simulations, and that measured values of the intercept parameter $b$ are difficult to reconcile with the existence of the host ISM.}
    \label{fig:corner}
\end{figure*}

The posterior for the 20 observed FRB sightlines, as well as the parameters best fitting the simulations, are shown in Fig.~\ref{fig:corner}. We tabulate best-fit values fitted parameters in Tab.~\ref{tab:params}. In Fig.~\ref{fig:camels}, the model prediction corresponding to the best-fit parameter values for the fiducial observational sample and fiducial simulation parameters are plotted as a set of three lines. In addition, 50 randomly-selected parameter draws from the posterior are shaded in translucent bands respectively.

We compare fitted values of $m,b,a,$ and $\sigma$ in the context of galaxy formation models, keeping in mind the slightly different interpretations of the fitted parameters in observational and simulation data. We refer to the fitted parameters for the observed and simulated sightlines using ``obs'' and ``sim'' subscripts respectively, referring to different simulation data by their model name if necessary. For all three models, Fig.~\ref{fig:corner} suggests $\sigma_{obs} > \sigma_{sim}$. This is consistent with observational data including contributions to the scatter from the host ISM and CBM, which are not included in the simulation predictions. 
Similarly, since $\langle \dmhism+\dmhcbm \rangle > 0$, we also expect that $b_{obs} > b_{sim}$. However, in the Astrid model, we see that $b_{obs} \approx b_{sim}$ to within $\pm 27$ pc cm$^{-3}$. 
This suggests that if the Astrid model accurately captures \dmhcgm, there is not much room for the host ISM, let alone the CBM, to contribute to the observed values of~\dmhm\citep{ocker2020electron}.

Finally, we see that $m_{obs} \neq m_\mathrm{Astrid}$ at the $4.2\sigma$ level; critically, the observations and simulations disagree not only in the magnitude but also the sign of the slope.
This discrepancy could potentially be explained by the different interpretation of $m_{obs}$ and $m_{sim}$, since $m_{obs}$ contains a host ISM contribution.
However, the discrepancy means that a large negative contribution to $m$ from the host ISM would be needed to reconcile our data with the Astrid prediction. We would need an anti-correlation between \dmhism\ and $M^*$ of $m \approx -180$ pc/cm$^{3}$/dex, i.e. that as FRB hosts have fewer stars they have thicker ionized gas disks. 

\subsection{Robustness of the results}\label{sec:robustness}
Our fiducial analysis provides two lines of evidence against the Astrid model.
One piece of evidence comes from the intercept parameter $b$. We observe that $b_{obs} \approx b_\mathrm{Astrid}$. 
However, due to the presence of the host ISM in the data and its deliberate omission from our simulation predictions, we would expect $b_{obs} > b_{sim}$. Another piece of evidence comes from the slope parameter $m$. We observe that $m_{obs} \neq m_\mathrm{Astrid}$ at the $4.2\sigma$ level, and that a strong negative correlation between \dmhism\ and $M^*$ would be needed to reconcile our data with the Astrid prediction for $m$.

We now vary our fiducial observational analysis and simulation analysis to explore how these two conclusions change, and explore whether the simulations can be reconciled with the data under different analysis choices.

The tension in the intercept parameter $b$ may be alleviated with several sources of systematics. Negatively-biased values of~\dmcosmic, negatively-biased values of $\langle \dmhism+\dmhcbm\rangle$, or values of~\dmmwcgm\ higher than predicted by~\citet{yamasaki2020galactic} could all alleviate the tension. In our ``$\langle\dmcosmic\rangle-30\%$'' analysis, we decrease the \dmcosmic\ values used in our \dmhm\ estimates. Table~\ref{tab:params} shows that this leaves some room for $\langle \dmhism+\dmhcbm\rangle = b_{obs} - b_{Astrid} \approx 14$ pc cm$^{-3}$.

Decreasing the Milky Way halo contribution may also alleviate the tension in $b$, since the Milky Way contributes to all sightlines. Averaged over all sightlines in our sample, $\langle\dmmwcgm\rangle = 37$ pc cm$^{-3}$~\citep{yamasaki2020galactic}, but extreme feedback scenarios could pull it lower than 30 pc cm$^{-3}$~\citep{keating2020exploring}, alleviating the tension by tens of units. If the Milky Way halo model also includes the gas beyond its virial radius as probed directly by the lowest FRB dispersion measures, a Milky Way contribution in the range 52-111 pc cm$^{-3}$~\citep{cook2023frb} would worsen the tension in $b$.

The tension in the slope parameter $m$ is more difficult to alleviate, since most systematics which may affect our measurements are correlated with either the stellar mass of the host or the DM of the FRB, but not both. Splitting the sample by redshift in half (``$z < 0.1$'' and ``$0.1 < z < 0.2$'' variants) suggests that the $z < 0.1$ half has a less steep slope, but the difference between the two halves is not statistically significant, and the reduction in the tension in $m$ can be somewhat attributed to halving the sample size: both halves have central parameter estimates of $m < 0$. 

One possibility is that optical selection effects preferentially associate FRBs with higher mass galaxies, though it is not obvious how this would affect the slope measurement. Nevertheless we check that tightening the association criteria used by the FRB association framework PATH~\citep[Probabilistically Associating Transients with their Hosts][]{aggarwal2021probabilistic} does not impact our conclusions. In the PATH formalism, $P(O|x)$ refers to the posterior probability that the FRB is associated with its a putative host; dropping observational data points with $P(O|x) < 0.95$ does not change our conclusions (see ``P(O|x) > 0.95'' variant in Table~\ref{tab:params}). 

We also test our sensitivity to $\sigmacosmic$. This is important because our conclusions about feedback in galaxies depend on our parameter estimates, which in turn may depend on feedback circularly via \sigmacosmic. For example, IllustrisTNG implements a feedback model stronger than Astrid and weaker than SIMBA, resulting in $\sigma_\textrm{cosmic} \approx 60$ pc cm$^{-3}$ at $z = 0$ and 105 pc cm$^{-3}$ at $z = 0.2$~\citep{walker2024dispersion}, whereas the original Illustris simulation finds $\approx15\%$ lower values for \sigmacosmic due to its more extreme feedback~\citealp[see Fig. 2 of]{jaroszynski2019fast}. Ideally, \sigmacosmic(z) could be measured from a large simulation volume at the same mass resolution employing the same sub-grid model in which the model predictions were generated. However, the small-volume CAMELS run of Astrid has a different mass resolution from the large-volume production run of Astrid and puts a direct calculation of $\sigmacosmic$ out of the scope of this work. We test our sensitivity to $\sigmacosmic$ by setting it to a constant independent of redshift (``\sigmacosmic = 80 pc cm$^{-3}$'' variant), finding that our parameter estimates are not sensitive to second-order effects of feedback through \sigmacosmic.

As an extreme test of robustness, we exclude the data point which most drives the slope measurement, FRB 20210807D. The statistical tension in $m$ values is reduced to $2.9\sigma$, but the slope ($m_{obs}$) remains negative. We regard this as an upper bound on the effect of single outliers in our sample, whose small size could be enlarged by upcoming FRB surveys~\citep{collaboration2025chimeoutriggers,wang2025craft,hallinan2019dsa,vanderlinde2019canadian}.

\begin{figure*}
\includegraphics[trim={0 2cm 0 2cm},clip,width=0.95\textwidth]{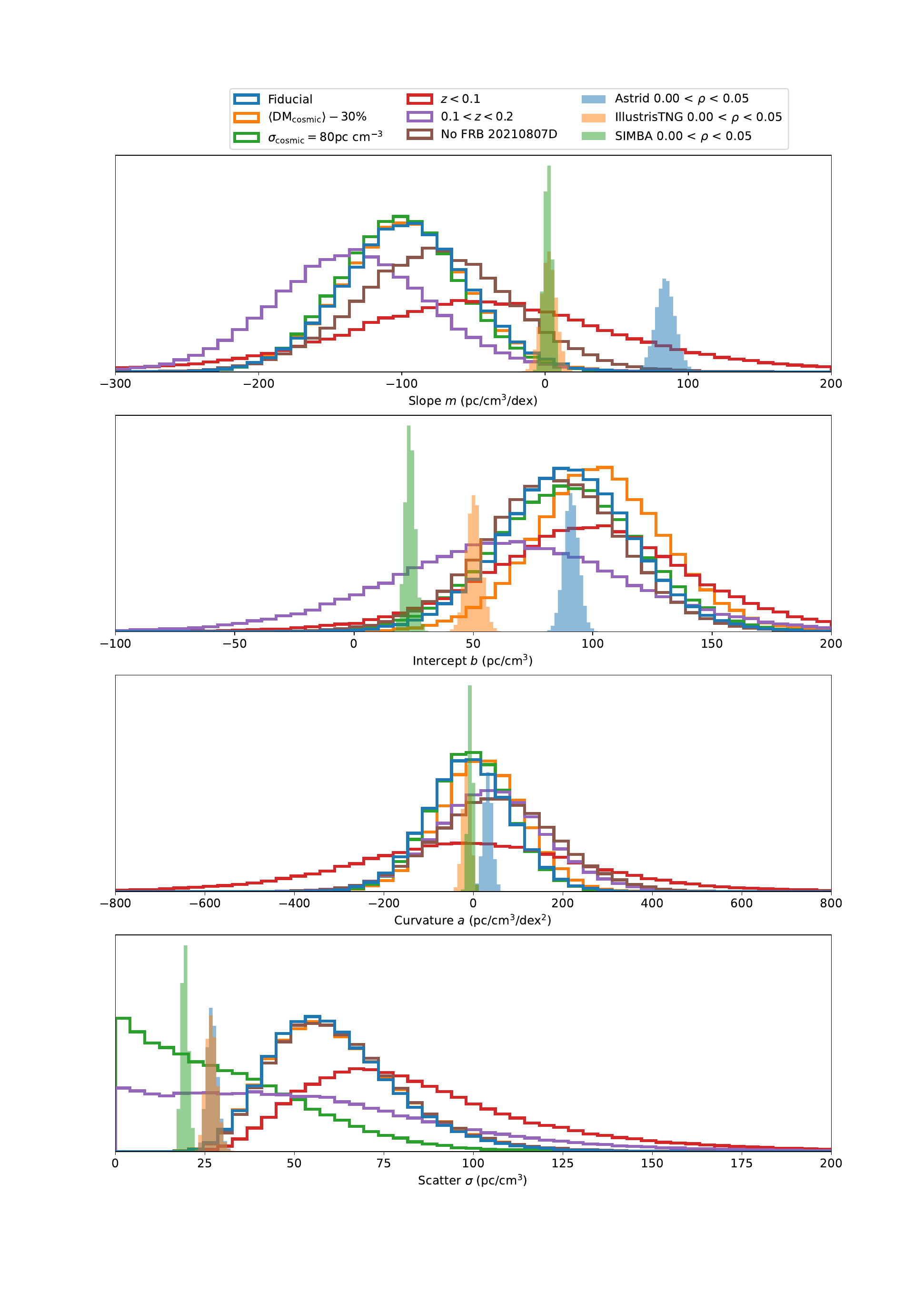}
\label{fig:params}
\caption{One-dimensional parameter estimates for $a,b,m,$ and $\sigma$ for our fiducial observational data (empty gray histogram), several variants thereof, as well as our simulation data (filled blue, orange, and green histograms). Parameter estimates are quoted in Tab.~\ref{tab:params}.}\end{figure*}
To check for the impact of varying the FRB host offset distribution, we measure $m$ in our simulations for different ranges by varying $\rho_\mathrm{min}$ and $\rho_\mathrm{max}$. Looking at $0.05 < \rho < 0.10$ and $0.10 < \rho < 0.15$ sightlines in the CAMELS predictions, finding that $m_\mathrm{Astrid}$ increases as FRBs get closer to their hosts. Since the maximum FRB host offset in our sample is $\rho_\mathrm{max} \lesssim 0.05$, it is likely that the  discrepancy between our observational data and the slope prediction of CAMELS-Astrid may be even larger than reported.

\section{Discussion}\label{sec:discussion}
Of the three fiducial simulations considered, our observations are most in tension with CAMELS-Astrid, while CAMELS-IllustrisTNG and CAMELS-SIMBA impose more modest requirements on the host ISM and CBM contributions to the slope $m$.
Unless the DM of the circum-burst medium is highly sensitive to global properties of the FRB host (e.g. its stellar mass), the discrepancy between the slope parameters measured in observations and simulations must be explained by the host ISM.
To resolve the discrepancy, our observations require that the host galaxy ISM can produce an anti-correlation between \dmhism and $M^*$. Astrid requires $m \approx 180$ pc/cm$^{-3}$/dex, whereas the stronger feedback feedback models implemented in IllustrisTNG and SIMBA require a more modest, but still negative, host ISM contribution of $m \approx -100$ pc/cm$^{-3}$/dex. 

Interestingly, a large contribution from the host ISM at low stellar mass $M^* \lesssim 10^{9}$ would be qualitatively consistent with the large \dmhm\ values measured in low-mass dwarf FRB hosts. These large values had previously been attributed at least in part to an extreme circum-burst medium on the basis of scattering timescales~\citep{caleb2023subarc}, persistent radio emission~\citep{marcote2020repeating,niu2021repeating,bhandari2023constraints}, and extreme rotation measures~\citep{michilli2018extreme}.

However, a strong negative correlation between stellar mass and \dmhism\ would be at odds with the positive correlation between $M^*$ and non-FRB based tracers of the ISM. For instance,~\citet{bernalescortes2025empirical} report a positive correlation between stellar mass and the component of \dmhism traced by H$\alpha$ emission measures in a sample of eleven FRB hosts with VLT/MUSE observations, suggesting a slope contribution of $m \approx 31$ pc/cm$^{3}$/dex (albeit over a slightly different range of host stellar masses). In addition, the mass of neutral hydrogen gas is known to be positively correlated with increasing stellar mass in HI-selected galaxies~\citep{huang2012arecibo,maddox2015variation,parkash2018relationships}.
The existence of positive correlations between ISM gas and stellar mass provide empirical evidence which suggest that the fiducial CAMELS-Astrid simulations should be disfavored. CAMELS-IllustrisTNG and CAMELS-SIMBA are more consistent with the data at the 2$\sigma$ level, but still require a host ISM contribution in contradiction with~\citet{bernalescortes2025empirical}. 

A detailed synthesis of these various correlations in terms of inflows and outflows between the ISM and CGM as a function of sub-grid model~\citep{wright2024baryon} is out of the scope of this work. 
Replicating the $M^*-\dmh$ relation measured in this paper and correlations such as those characterized by~\citet{bernalescortes2025empirical} in high-resolution simulations including realistic radiative transfer~\citep[such as extensions of][]{orr2024objects} will be crucial to simultaneously understanding the impact of the host ISM and CGM on the observed DM of FRBs. 

Furthermore, it is an open question whether stellar or AGN feedback is principally responsible for feedback effects in isolated galaxies. In this case, further analysis of the CAMELS suite of simulations could explore the divergence between observations and simulations in the slope of the $M^*-\dmh$ relation. Hydrodynamic simulations of stellar and AGN feedback in isolated galaxies~\citep{newton2013study} suggest that stellar feedback dominates in isolated galaxies in the absence of galaxy merger activity. This leads to the notion that in this mass range, we are primarily sensitive to stellar feedback. 
The negative slope exhibited by our data may be interpreted in terms of the cold flow paradigm of galaxy formation, which suggests that cool gas flows in sufficiently-massive halos $\gtrsim 10^{12} M_\odot$~\citep[see e.g.,][]{dekel2006galaxy} are shock-heated and rarefied, making them vulnerable to expulsion by \emph{either} stellar or AGN feedback effects. 

On the other hand, as pointed out by~\citet{ni2023camels}, the principal difference that sets apart Astrid from IllustrisTNG and SIMBA is the kinetic AGN feedback in ASTRID, which is activated in a more stringent way and with a lower maximum efficiency than in the IllustrisTNG model. In addition, it is known that prescriptions for black hole seeding and dynamics differ and are potentially less realistic in the lower resolution Astrid runs in CAMELS (referred to in~\citep{ni2023camels} as CAMELS-Astrid) than in the production run of Astrid. SIMBA, which predicts the lowest \dmhcgm, notably features anisotropic AGN feedback~\citep{anglesalcazar2017black,dave2019simba,khrykin2024cosmic} in which collimated jets expel gas from their halos, and it is thought that anisotropic feedback more effectively expels gas to larger radii than isotropic feedback. 

It is possible that both stellar \emph{and} AGN feedback play a role in this mass range, since the reservoir of ionized gas in the CGM couples to both stellar and AGN feedback processes. 
This causes a complex interplay between star formation and AGN activity~\citep{2025ApJ...980...61M}. 
For instance, at low black hole masses and high redshift, stellar feedback regulates the growth of supermassive black holes by limiting their accretion rate~\citep{anglesalcazar2017black}. 
In addition, ~\citet{booth2013interaction} suggests, contrary to~\citet{newton2013study}, that stellar and AGN feedback have opposing influences on the star formation rate: it is plausible that similar competing effects may shape the $M^*-\dmh$ relation. 
Interestingly, the data point with the lowest value of \dmhm\ and highest stellar mass (FRB 20210807A) corresponds to a galaxy which exhibits both a high star formation rate and AGN activity~\citep{gordon2023demographics}. 
Omitting that data point from the fit (``No FRB 20210807D'' variant in Fig.~\ref{fig:params}) somewhat alleviates the statistical tension in $m$ values between observations and simulations to $\approx 3$ sigma, but does not change our conclusion that \dmhism\ contribution needs to have a negative slope with stellar mass to reconcile any of the three CAMELS simulations with the data.

\begin{figure}
    \centering
    \includegraphics[width=\linewidth]{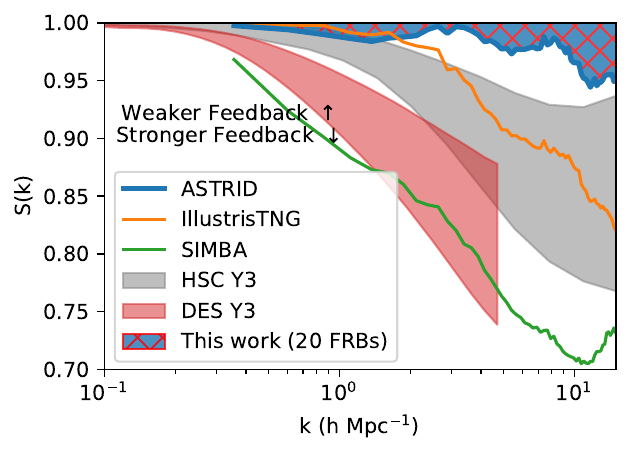}
    \caption{The level of suppression of the matter power spectrum on small scales due to baryonic feedback as predictions from ASTRID, IllustrisTNG, and SIMBA, as well as the official DES Y3 and HSC analyses~\citep{chen2023constraining,terasawa2025exploring}. The Astrid model, which corresponds to weak feedback, is difficult to reconcile with observations, implying an upper limit on $S(k,z)$.}
    \label{fig:sk}
\end{figure}

As pointed out by~\citet{medlock2025constraining}, our conclusion about the strength of feedback can be compared with other measurements of baryonic feedback from weak lensing, albeit indirectly. 
We quantify the strength of feedback using the matter power spectrum suppression factor. 
In simulations like CAMELS, this is defined as $S(k) \equiv P_{\rm hydro}/P_{\rm nbody}$ calculated from the same snapshots used in \S~\ref{sec:camels} for our $M^*-\dmh$ measurements. The suppression factors measured in the same CAMELS snapshots are plotted in Fig.~\ref{fig:sk} for the three models considered here, where $S(k) = 1$ corresponds to no feedback (i.e. gas traces dark matter exactly), and $S(k) \to 0$ corresponds to strong feedback which causes a suppression of the matter power spectrum.
The suppression factor inferred from Dark Energy Survey~\citep{chen2023constraining} and Subaru Hyper Suprime Cam~\citep{terasawa2025exploring} power spectra are also plotted for comparison in Fig.~\ref{fig:sk}.

While our conclusions are difficult to compare directly with other gas probes, our conclusion is qualitatively consistent with findings from~\citet{hadzhiyska2024evidence}, which finds that the feedback implementation in IllustrisTNG is too weak to explain observations of the kinetic Sunyaev-Zeldovich effect. We can provide a comparison to other measurements of baryonic feedback subject to the important caveat that the small volume of CAMELS snapshots considered in this work does not sample the redshift range or the range of the halo mass function to which cosmic shear measurements are most sensitive~\citep{tr_oster2022joint,aric_o2023des,terasawa2025exploring}. Our comparison probes feedback from galaxies in the range $10^9 < M/M_\odot < 10^{11}$ which correspond roughly to halos at $M_h \sim 10^{11-13} M_\odot$ halos at $z \sim 0$. In contrast, cosmic shear analyses target halo masses of $M_h \sim 10^{12-14}$ at earlier times ($z \sim 0.5$ for the Kilo-Degree Survey and DES and $z\sim 0.9$ for Subaru-HSC). Since CAMELS does not capture structure on scales larger than the box size of 25 Mpc/h ($k > 0.03$ Mpc$^{-1}$), it does not sample the full range of the halo mass function, or include contributions to the smoothing of the density field on large scales from group and cluster-scale halos.
We thus consider our results as a lower limit on the impact of feedback: in other words, an indirect upper limit on $S(k,z=0)$. We note that other approaches, short of direct angular cross-correlations~\citep{wang2025measurement}, also quantify feedback through metrics other than $S(k,z)$. Those focus on quantities like the DM-redshift distribution~\citep{macquart2020census,james2021z,sharma2025hydrodynamical}, the baryon spread~\citep{medlock2025constraining} and IGM/CGM baryon fractions~\citep[see e.g.,][]{khrykin2024flimflam,khrykin2024cosmic,connor2024gas,hussaini2025correlation}, and like the work presented here, rely on simulation snapshots to connect their inferences of feedback to $S(k,z)$.

\section{Conclusions and Future Work}
In this paper, we have pointed out that the relationship between the stellar mass of a galaxy and its \dmhm\ is a sensitive probe of baryonic feedback in isolated galaxies. Our findings are as follows:
\begin{enumerate}
    \item We have used simulations and a sample of FRB data to explore the $M^*-\dmh$ relation for the first time. We have characterized it in terms of its slope, intercept, curvature, and its scatter. In models of weak feedback like the fiducial Astrid model in CAMELS, as $M^*$ increases, so does \dmhcgm. 
    \item The observations show -- contrary to weak feedback models -- that \dmhm tends to be weakly \emph{anti-}correlated with stellar mass in FRB host galaxies in the mass range $10^{9-11} M_\odot$ at $z \sim 0.1$. 
    \item In particular, both the slope and intercept measurements from Astrid are hard to explain without invoking a negative correlation between stellar mass and \dmhism. Self-consistent, high-resolution modeling of the host ISM and CGM are needed to rigorously assess whether this is plausible, but observations indicate that some ISM observables (the HI mass and the H$\alpha$-inferred DM) are positively correlated with stellar mass, providing some empirical evidence against this possibility.
    \item Our measurements at present are in conflict with all three feedback models considered, with the predictions of Astrid being the most discrepant. They constrain the gas environments of FRB host galaxies at $z = 0$ and $10^9 < M^*/M_\odot < 10^{11}$, and imply an indirect lower limit on the strength of baryonic feedback via the suppression factor of the matter power spectrum $S(k,z=0)$ due to $M_h \lesssim 10^{13} M_\odot$ halos.
\end{enumerate}

This analysis could readily be extended in several ways. Going to larger volumes (higher redshifts) better samples the halo mass function using FRB hosts living in groups and clusters like those presented by~\citet{connor2023deep} and~\citet{chime2025catalog}. 
Well-matched samples as a function of redshift could be used to study the redshift evolution of the gas environments of galaxies over cosmic time. 
However, the foreground posed by the IGM will demand a larger number of sources, since $\sigmacosmic$ increases with increasing redshift. 
In the opposite direction, widefield telescopes like CHORD~\citep{vanderlinde2019canadian} and coherent all-sky monitors like BURSTT~\citep{lin2022burstt,connor2022stellar} may soon deliver an order-of-magnitude increase in FRBs closer than $<$ 400 Mpc. 
Doing so will extend the $M^*-\dmh$ relation downwards in stellar and halo mass, which can allow baryonic feedback to be disentangled from dark matter effects in dark-matter dominated systems like dwarf galaxies. In the long run this could open up the new sub-field of ``near-field FRB cosmology''. 

\begin{table*}[htbp]
\centering
\begin{tabular}{|c|c|c|c|c|}
\hline
Model & $m$ & $b$ & $a$ & $\sigma$ \\
\hline
Fiducial & $-97^{+44}_{-44}$ & $89^{+27}_{-27}$ & $-11^{+94}_{-93}$ & $59^{+20}_{-15}$ \\
$\langle \mathrm{DM}_\mathrm{cosmic}\rangle - 30\%$ & $-98^{+43}_{-44}$ & $103^{+28}_{-27}$ & $18^{+94}_{-93}$ & $60^{+21}_{-16}$ \\
$\sigma_\mathrm{cosmic} = 80 \mathrm{pc~cm}^{-3}$ & $-101^{+41}_{-42}$ & $89^{+30}_{-30}$ & $-7^{+88}_{-89}$ & $26^{+28}_{-19}$ \\
$z < 0.1$ & $-43^{+97}_{-92}$ & $99^{+44}_{-44}$ & $-12^{+261}_{-263}$ & $79^{+37}_{-24}$ \\
$0.1 < z < 0.2$ & $-134^{+55}_{-55}$ & $64^{+51}_{-50}$ & $39^{+124}_{-123}$ & $48^{+43}_{-33}$ \\
No FRB 20210807D & $-77^{+54}_{-54}$ & $84^{+28}_{-29}$ & $50^{+134}_{-131}$ & $60^{+20}_{-16}$ \\
P(O|x) > $0.95$ & $-104^{+40}_{-41}$ & $110^{+26}_{-27}$ & $-42^{+90}_{-90}$ & $48^{+19}_{-15}$ \\
$\sigma_\mathrm{cosmic}$ + 50\% & $-102^{+53}_{-52}$ & $90^{+29}_{-29}$ & $-12^{+111}_{-112}$ & $51^{+22}_{-17}$ \\
\hline
Astrid 0.00 < $\rho$ < 0.05 & $84.0^{+6.9}_{-6.8}$ & $91.0^{+3.1}_{-3.1}$ & $32.0^{+9.8}_{-9.9}$ & $27.0^{+1.6}_{-1.4}$ \\
Astrid 0.05 < $\rho$ < 0.10 & $55.6^{+3.2}_{-3.2}$ & $49.7^{+1.4}_{-1.3}$ & $27.3^{+4.2}_{-4.1}$ & $13.8^{+0.6}_{-0.6}$ \\
Astrid 0.10 < $\rho$ < 0.15 & $29.8^{+1.7}_{-1.7}$ & $25.2^{+0.8}_{-0.8}$ & $14.0^{+2.3}_{-2.5}$ & $6.7^{+0.3}_{-0.4}$ \\
Astrid 0.15 < $\rho$ < 0.20 & $10.2^{+0.7}_{-0.8}$ & $10.1^{+0.3}_{-0.3}$ & $4.5^{+1.0}_{-1.1}$ & $2.7^{+0.1}_{-0.2}$ \\
IllustrisTNG 0.00 < $\rho$ < 0.05 & $2.5^{+5.3}_{-5.3}$ & $50.2^{+3.1}_{-3.1}$ & $-15.3^{+9.3}_{-9.4}$ & $26.8^{+1.7}_{-1.5}$ \\
IllustrisTNG 0.05 < $\rho$ < 0.10 & $1.0^{+2.1}_{-1.9}$ & $25.8^{+1.1}_{-1.1}$ & $-9.8^{+3.4}_{-3.4}$ & $10.8^{+0.5}_{-0.5}$ \\
IllustrisTNG 0.10 < $\rho$ < 0.15 & $2.0^{+1.6}_{-1.5}$ & $13.9^{+0.9}_{-0.9}$ & $-4.4^{+2.5}_{-2.6}$ & $8.6^{+0.4}_{-0.4}$ \\
IllustrisTNG 0.15 < $\rho$ < 0.20 & $0.0^{+0.6}_{-0.5}$ & $5.4^{+0.3}_{-0.3}$ & $-1.4^{+0.9}_{-0.9}$ & $2.8^{+0.1}_{-0.1}$ \\
SIMBA 0.00 < $\rho$ < 0.05 & $1.9^{+2.9}_{-2.9}$ & $23.4^{+2.1}_{-2.0}$ & $-6.7^{+5.2}_{-5.3}$ & $19.4^{+1.0}_{-0.9}$ \\
SIMBA 0.05 < $\rho$ < 0.10 & $-0.6^{+0.8}_{-0.8}$ & $11.9^{+0.5}_{-0.5}$ & $-5.1^{+1.3}_{-1.3}$ & $6.2^{+0.3}_{-0.2}$ \\
SIMBA 0.10 < $\rho$ < 0.15 & $0.1^{+0.5}_{-0.5}$ & $5.9^{+0.3}_{-0.3}$ & $-2.3^{+0.8}_{-0.8}$ & $3.2^{+0.1}_{-0.1}$ \\
SIMBA 0.15 < $\rho$ < 0.20 & $-0.1^{+0.2}_{-0.2}$ & $2.5^{+0.1}_{-0.1}$ & $-1.3^{+0.3}_{-0.4}$ & $1.5^{+0.1}_{-0.1}$ \\
\hline
\end{tabular}
\caption{MCMC parameter estimates for observed and simulated sightlines, where we fit the slope ($m$), intercept ($b$), curvature ($a$), and scatter ($\sigma$). Parameters estimates from the data plotted in Fig. 3 correspond to the “Fiducial” observational sample parameters. Other variants of our analysis of the observational data are shown in the upper half of the table. The lower half of the table shows the results of our simulation analysis using the fiducial impact parameter choice (``0.00 < $\rho$ < 0.05'') and several variations on it. Our measurement of $m$ in the data is difficult to reconcile with the $m$ values inferred from CAMELS-Astrid.}
\label{tab:params}
\end{table*}

\section{Acknowledgments}
We acknowledge that CHIME and~\kkoname\ are located on the traditional, 
ancestral, and unceded territory of the Syilx/Okanagan people. We are grateful to the staff of the Dominion Radio
Astrophysical Observatory, which is operated by the National Research Council of Canada. 
CHIME is funded by a grant from the Canada Foundation 
for Innovation (CFI) Leading Edge Fund (Project 31170) 
and by contributions from the provinces of British Columbia, 
Qu\'ebec and Ontario. The CHIME/FRB Project is funded by a 
grant from the CFI 2015 Innovation Fund (Project 33213) and 
by contributions from the provinces of British Columbia and 
Qu\'ebec, and by the Dunlap Institute for Astronomy and 
Astrophysics at the University of Toronto. 
Additional support was provided by the Canadian 
Institute for Advanced Research (CIFAR), McGill 
University and the McGill Space Institute via the 
Trottier Family Foundation, and the University of 
British Columbia. 
The CHIME/FRB Outriggers program is funded by 
the Gordon and Betty Moore Foundation and by a National Science Foundation (NSF) grant (2008031).
The \kkoname~Outrigger is situated on land leased from the Imperial Metals Corporation
FRB research at MIT is supported by an NSF grant (2008031).
FRB research at WVU is supported by an NSF grant (2006548, 2018490).

The Pan-STARRS1 Surveys (PS1) and the PS1 public science archive have been made possible through contributions by the Institute for Astronomy, the University of Hawaii, the Pan-STARRS Project Office, the Max-Planck Society and its participating institutes, the Max Planck Institute for Astronomy, Heidelberg and the Max Planck Institute for Extraterrestrial Physics, Garching, The Johns Hopkins University, Durham University, the University of Edinburgh, the Queen's University Belfast, the Harvard-Smithsonian Center for Astrophysics, the Las Cumbres Observatory Global Telescope Network Incorporated, the National Central University of Taiwan, the Space Telescope Science Institute, the National Aeronautics and Space Administration under Grant No. NNX08AR22G issued through the Planetary Science Division of the NASA Science Mission Directorate, the National Science Foundation Grant No. AST-1238877, the University of Maryland, Eotvos Lorand University (ELTE), the Los Alamos National Laboratory, and the Gordon and Betty Moore Foundation.

The Legacy Surveys consist of three individual and complementary projects: the Dark Energy Camera Legacy Survey (DECaLS; Proposal ID \#2014B-0404; PIs: David Schlegel and Arjun Dey), the Beijing-Arizona Sky Survey (BASS; NOAO Prop. ID \#2015A-0801; PIs: Zhou Xu and Xiaohui Fan), and the Mayall z-band Legacy Survey (MzLS; Prop. ID \#2016A-0453; PI: Arjun Dey). DECaLS, BASS and MzLS together include data obtained, respectively, at the Blanco telescope, Cerro Tololo Inter-American Observatory, NSF’s NOIRLab; the Bok telescope, Steward Observatory, University of Arizona; and the Mayall telescope, Kitt Peak National Observatory, NOIRLab. 

Pipeline processing and analyses of the data were supported by NOIRLab and the Lawrence Berkeley National Laboratory (LBNL). The Legacy Surveys project is honored to be permitted to conduct astronomical research on Iolkam Du’ag (Kitt Peak), a mountain with particular significance to the Tohono O’odham Nation.

NOIRLab is operated by the Association of Universities for Research in Astronomy (AURA) under a cooperative agreement with the National Science Foundation. LBNL is managed by the Regents of the University of California under contract to the U.S. Department of Energy.

This project used data obtained with the Dark Energy Camera (DECam), which was constructed by the Dark Energy Survey (DES) collaboration. Funding for the DES Projects has been provided by the U.S. Department of Energy, the U.S. National Science Foundation, the Ministry of Science and Education of Spain, the Science and Technology Facilities Council of the United Kingdom, the Higher Education Funding Council for England, the National Center for Supercomputing Applications at the University of Illinois at Urbana-Champaign, the Kavli Institute of Cosmological Physics at the University of Chicago, Center for Cosmology and Astro-Particle Physics at the Ohio State University, the Mitchell Institute for Fundamental Physics and Astronomy at Texas A\&M University, Financiadora de Estudos e Projetos, Fundacao Carlos Chagas Filho de Amparo, Financiadora de Estudos e Projetos, Fundacao Carlos Chagas Filho de Amparo a Pesquisa do Estado do Rio de Janeiro, Conselho Nacional de Desenvolvimento Cientifico e Tecnologico and the Ministerio da Ciencia, Tecnologia e Inovacao, the Deutsche Forschungsgemeinschaft and the Collaborating Institutions in the Dark Energy Survey. The Collaborating Institutions are Argonne National Laboratory, the University of California at Santa Cruz, the University of Cambridge, Centro de Investigaciones Energeticas, Medioambientales y Tecnologicas-Madrid, the University of Chicago, University College London, the DES-Brazil Consortium, the University of Edinburgh, the Eidgenossische Technische Hochschule (ETH) Zurich, Fermi National Accelerator Laboratory, the University of Illinois at Urbana-Champaign, the Institut de Ciencies de l’Espai (IEEC/CSIC), the Institut de Fisica d’Altes Energies, Lawrence Berkeley National Laboratory, the Ludwig Maximilians Universitat Munchen and the associated Excellence Cluster Universe, the University of Michigan, NSF’s NOIRLab, the University of Nottingham, the Ohio State University, the University of Pennsylvania, the University of Portsmouth, SLAC National Accelerator Laboratory, Stanford University, the University of Sussex, and Texas A\&M University.

BASS is a key project of the Telescope Access Program (TAP), which has been funded by the National Astronomical Observatories of China, the Chinese Academy of Sciences (the Strategic Priority Research Program “The Emergence of Cosmological Structures” Grant \# XDB09000000), and the Special Fund for Astronomy from the Ministry of Finance. The BASS is also supported by the External Cooperation Program of Chinese Academy of Sciences (Grant \# 114A11KYSB20160057), and Chinese National Natural Science Foundation (Grant \# 12120101003, \# 11433005).

The Legacy Survey team makes use of data products from the Near-Earth Object Wide-field Infrared Survey Explorer (NEOWISE), which is a project of the Jet Propulsion Laboratory/California Institute of Technology. NEOWISE is funded by the National Aeronautics and Space Administration.

The Legacy Surveys imaging of the DESI footprint is supported by the Director, Office of Science, Office of High Energy Physics of the U.S. Department of Energy under Contract No. DE-AC02-05CH1123, by the National Energy Research Scientific Computing Center, a DOE Office of Science User Facility under the same contract; and by the U.S. National Science Foundation, Division of Astronomical Sciences under Contract No. AST-0950945 to NOAO.

The Photometric Redshifts for the Legacy Surveys (PRLS) catalog used in this paper was produced thanks to funding from the U.S. Department of Energy Office of Science, Office of High Energy Physics via grant DE-SC0007914.

This work has made use of data from the European Space Agency (ESA) mission
{\it Gaia} (\url{https://www.cosmos.esa.int/gaia}), processed by the {\it Gaia}
Data Processing and Analysis Consortium (DPAC,
\url{https://www.cosmos.esa.int/web/gaia/dpac/consortium}). Funding for the DPAC
has been provided by national institutions, in particular the institutions
participating in the {\it Gaia} Multilateral Agreement.

This research has made use of the CIRADA cutout service (\url{cutouts.cirada.ca}) operated by the Canadian Initiative for Radio Astronomy Data Analysis (CIRADA). CIRADA is funded by a grant from the Canada Foundation for Innovation 2017 Innovation Fund (Project 35999), as well as by the Provinces of Ontario, British Columbia, Alberta, Manitoba and Quebec, in collaboration with the National Research Council of Canada, the US National Radio Astronomy Observatory and Australia’s Commonwealth Scientific and Industrial Research Organisation. 

We would like to thank Alexa Gordon for providing inclination angle measurements for the host galaxies of FRB 20240201A and 20240310A. This work is dedicated to the memory of Thomas Leung (1934-2025).

\vspace{5mm}
\facilities{CHIME}
\software{\texttt{numpy} \citep{harris2020array}, \texttt{scipy} \citep{virtanen2020scipy}, \texttt{matplotlib}~\citep{hunter2007matplotlib}, 
\texttt{PyFX},\texttt{coda}~\citet{leung2024vlbi},\texttt{difxcalc}~\citet{gordon2016difxcalc}
}
\bibliographystyle{aasjournal}
\bibliography{references}

\begin{thebibliography}{}
\expandafter\ifx\csname natexlab\endcsname\relax\def\natexlab#1{#1}\fi
\providecommand{\url}[1]{\href{#1}{#1}}
\providecommand{\dodoi}[1]{doi:~\href{http://doi.org/#1}{\nolinkurl{#1}}}
\providecommand{\doeprint}[1]{\href{http://ascl.net/#1}{\nolinkurl{http://ascl.net/#1}}}
\providecommand{\doarXiv}[1]{\href{https://arxiv.org/abs/#1}{\nolinkurl{https://arxiv.org/abs/#1}}}

\bibitem[{{Aggarwal} {et~al.}(2021){Aggarwal}, {Budav{\'a}ri}, {Deller}, {Eftekhari}, {James}, {Prochaska}, \& {Tendulkar}}]{aggarwal2021probabilistic}
{Aggarwal}, K., {Budav{\'a}ri}, T., {Deller}, A.~T., {et~al.} 2021, \apj, 911, 95, \dodoi{10.3847/1538-4357/abe8d2}

\bibitem[{{Angl{\'e}s-Alc{\'a}zar} {et~al.}(2017){Angl{\'e}s-Alc{\'a}zar}, {Faucher-Gigu{\`e}re}, {Quataert}, {Hopkins}, {Feldmann}, {Torrey}, {Wetzel}, \& {Kere{\v{s}}}}]{anglesalcazar2017black}
{Angl{\'e}s-Alc{\'a}zar}, D., {Faucher-Gigu{\`e}re}, C.-A., {Quataert}, E., {et~al.} 2017, \mnras, 472, L109, \dodoi{10.1093/mnrasl/slx161}

\bibitem[{{Aric{\`o}} {et~al.}(2023){Aric{\`o}}, {Angulo}, {Zennaro}, {Contreras}, {Chen}, \& {Hern{\'a}ndez-Monteagudo}}]{aric_o2023des}
{Aric{\`o}}, G., {Angulo}, R.~E., {Zennaro}, M., {et~al.} 2023, \aap, 678, A109, \dodoi{10.1051/0004-6361/202346539}

\bibitem[{{Bahk} \& {Hwang}(2024)}]{bahk2024galaxy}
{Bahk}, H., \& {Hwang}, H.~S. 2024, \apjs, 272, 7, \dodoi{10.3847/1538-4365/ad323f}

\bibitem[{{Battaglia} {et~al.}(2012){Battaglia}, {Bond}, {Pfrommer}, \& {Sievers}}]{battaglia2012on}
{Battaglia}, N., {Bond}, J.~R., {Pfrommer}, C., \& {Sievers}, J.~L. 2012, \apj, 758, 75, \dodoi{10.1088/0004-637X/758/2/75}

\bibitem[{{Batten} {et~al.}(2021){Batten}, {Duffy}, {Wijers}, {Gupta}, {Flynn}, {Schaye}, \& {Ryan-Weber}}]{batten2021cosmic}
{Batten}, A.~J., {Duffy}, A.~R., {Wijers}, N.~A., {et~al.} 2021, \mnras, 505, 5356, \dodoi{10.1093/mnras/stab1528}

\bibitem[{{Bernales-Cortes} {et~al.}(2025){Bernales-Cortes}, {Tejos}, {Prochaska}, {Khrykin}, {Marnoch}, {Ryder}, \& {Shannon}}]{bernalescortes2025empirical}
{Bernales-Cortes}, L., {Tejos}, N., {Prochaska}, J.~X., {et~al.} 2025, \aap, 696, A81, \dodoi{10.1051/0004-6361/202452026}

\bibitem[{{Bhandari} {et~al.}(2020){Bhandari}, {Sadler}, {Prochaska}, {Simha}, {Ryder}, {Marnoch}, {Bannister}, {Macquart}, {Flynn}, {Shannon}, {Tejos}, {Corro-Guerra}, {Day}, {Deller}, {Ekers}, {Lopez}, {Mahony}, {Nu{\~n}ez}, \& {Phillips}}]{bhandari2020host}
{Bhandari}, S., {Sadler}, E.~M., {Prochaska}, J.~X., {et~al.} 2020, \apjl, 895, L37, \dodoi{10.3847/2041-8213/ab672e}

\bibitem[{{Bhandari} {et~al.}(2023{\natexlab{a}}){Bhandari}, {Gordon}, {Scott}, {Marnoch}, {Sridhar}, {Kumar}, {James}, {Qiu}, {Bannister}, {T. Deller}, {Eftekhari}, {Fong}, {Glowacki}, {Prochaska}, {Ryder}, {Shannon}, \& {Simha}}]{bhandari2023nonrepeating}
{Bhandari}, S., {Gordon}, A.~C., {Scott}, D.~R., {et~al.} 2023{\natexlab{a}}, \apj, 948, 67, \dodoi{10.3847/1538-4357/acc178}

\bibitem[{{Bhandari} {et~al.}(2023{\natexlab{b}}){Bhandari}, {Marcote}, {Sridhar}, {Eftekhari}, {Hessels}, {Hewitt}, {Kirsten}, {Ould-Boukattine}, {Paragi}, \& {Snelders}}]{bhandari2023constraints}
{Bhandari}, S., {Marcote}, B., {Sridhar}, N., {et~al.} 2023{\natexlab{b}}, \apjl, 958, L19, \dodoi{10.3847/2041-8213/ad083f}

\bibitem[{{Bhardwaj} {et~al.}(2024{\natexlab{a}}){Bhardwaj}, {Lee}, \& {Ji}}]{bhardwaj2024selection}
{Bhardwaj}, M., {Lee}, J., \& {Ji}, K. 2024{\natexlab{a}}, arXiv e-prints, arXiv:2408.01876, \dodoi{10.48550/arXiv.2408.01876}

\bibitem[{{Bhardwaj} {et~al.}(2021{\natexlab{a}}){Bhardwaj}, {Gaensler}, {Kaspi}, {Landecker}, {Mckinven}, {Michilli}, {Pleunis}, {Tendulkar}, {Andersen}, {Boyle}, {Cassanelli}, {Chawla}, {Cook}, {Dobbs}, {Fonseca}, {Kaczmarek}, {Leung}, {Masui}, {Mnchmeyer}, {Ng}, {Rafiei-Ravandi}, {Scholz}, {Shin}, {Smith}, {Stairs}, \& {Zwaniga}}]{bhardwaj2021nearby}
{Bhardwaj}, M., {Gaensler}, B.~M., {Kaspi}, V.~M., {et~al.} 2021{\natexlab{a}}, \apjl, 910, L18, \dodoi{10.3847/2041-8213/abeaa6}

\bibitem[{{Bhardwaj} {et~al.}(2021{\natexlab{b}}){Bhardwaj}, {Kirichenko}, {Michilli}, {Mayya}, {Kaspi}, {Gaensler}, {Rahman}, {Tendulkar}, {Fonseca}, {Josephy}, {Leung}, {Merryfield}, {Petroff}, {Pleunis}, {Sanghavi}, {Scholz}, {Shin}, {Smith}, \& {Stairs}}]{bhardwaj2021local}
{Bhardwaj}, M., {Kirichenko}, A.~Y., {Michilli}, D., {et~al.} 2021{\natexlab{b}}, arXiv e-prints, arXiv:2108.12122.
\newblock \doarXiv{2108.12122}

\bibitem[{{Bhardwaj} {et~al.}(2024{\natexlab{b}}){Bhardwaj}, {Michilli}, {Kirichenko}, {Modilim}, {Shin}, {Kaspi}, {Andersen}, {Cassanelli}, {Brar}, {Chatterjee}, {Cook}, {Dong}, {Fonseca}, {Gaensler}, {Ibik}, {Kaczmarek}, {Lanman}, {Leung}, {Masui}, {Pandhi}, {Pearlman}, {Petroff}, {Pleunis}, {Prochaska}, {Rafiei-Ravandi}, {Sand}, {Scholz}, \& {Smith}}]{bhardwaj2024host}
{Bhardwaj}, M., {Michilli}, D., {Kirichenko}, A.~Y., {et~al.} 2024{\natexlab{b}}, \apjl, 971, L51, \dodoi{10.3847/2041-8213/ad64d1}

\bibitem[{{Booth} \& {Schaye}(2013)}]{booth2013interaction}
{Booth}, C.~M., \& {Schaye}, J. 2013, Scientific Reports, 3, 1738, \dodoi{10.1038/srep01738}

\bibitem[{{Boquien} {et~al.}(2019){Boquien}, {Burgarella}, {Roehlly}, {Buat}, {Ciesla}, {Corre}, {Inoue}, \& {Salas}}]{boquien2019cigale}
{Boquien}, M., {Burgarella}, D., {Roehlly}, Y., {et~al.} 2019, \aap, 622, A103, \dodoi{10.1051/0004-6361/201834156}

\bibitem[{{Caleb} {et~al.}(2023){Caleb}, {Driessen}, {Gordon}, {Tejos}, {Bernales}, {Qiu}, {Chibueze}, {Stappers}, {Rajwade}, {Cavallaro}, {Wang}, {Kumar}, {Majid}, {Wharton}, {Naudet}, {Bezuidenhout}, {Jankowski}, {Malenta}, {Morello}, {Sanidas}, {Surnis}, {Barr}, {Chen}, {Kramer}, {Fong}, {Kilpatrick}, {Prochaska}, {Simha}, {Venter}, {Heywood}, {Kundu}, \& {Schussler}}]{caleb2023subarc}
{Caleb}, M., {Driessen}, L.~N., {Gordon}, A.~C., {et~al.} 2023, \mnras, 524, 2064, \dodoi{10.1093/mnras/stad1839}

\bibitem[{{Cassanelli} {et~al.}(2024){Cassanelli}, {Leung}, {Sanghavi}, {Mena-Parra}, {Cary}, {Mckinven}, {Bhardwaj}, {Masui}, {Michilli}, {Bandura}, {Chatterjee}, {Peterson}, {Kaczmarek}, {Patel}, {Rahman}, {Shin}, {Vanderlinde}, {Berger}, {Brar}, {Boyle}, {Breitman}, {Chawla}, {Curtin}, {Dobbs}, {Dong}, {Fonseca}, {Gaensler}, {Ibik}, {Kaspi}, {Kholoud}, {Lanman}, {Lazda}, {Lin}, {Luo}, {Meyers}, {Milutinovic}, {Ng}, {Noble}, {Pearlman}, {Pen}, {Petroff}, {Pleunis}, {Quine}, {Rafiei-Ravandi}, {Renard}, {Sand}, {Schoen}, {Scholz}, {Smith}, {Stairs}, \& {Tendulkar}}]{cassanelli2024fast}
{Cassanelli}, T., {Leung}, C., {Sanghavi}, P., {et~al.} 2024, Nature Astronomy, 1, \dodoi{10.48550/arXiv.2307.09502}

\bibitem[{{Chabrier}(2003)}]{chabrier2003galactic}
{Chabrier}, G. 2003, \pasp, 115, 763, \dodoi{10.1086/376392}

\bibitem[{{Chen} {et~al.}(2023){Chen}, {Aric{\`o}}, {Huterer}, {Angulo}, {Weaverdyck}, {Friedrich}, {Secco}, {Hern{\'a}ndez-Monteagudo}, {Alarcon}, {Alves}, {Amon}, {Andrade-Oliveira}, {Baxter}, {Bechtol}, {Becker}, {Bernstein}, {Blazek}, {Brandao-Souza}, {Bridle}, {Camacho}, {Campos}, {Carnero Rosell}, {Carrasco Kind}, {Cawthon}, {Chang}, {Chen}, {Chintalapati}, {Choi}, {Cordero}, {Crocce}, {Pereira}, {Davis}, {DeRose}, {Di Valentino}, {Diehl}, {Dodelson}, {Doux}, {Drlica-Wagner}, {Eckert}, {Eifler}, {Elsner}, {Elvin-Poole}, {Everett}, {Fang}, {Fert{\'e}}, {Fosalba}, {Gatti}, {Gaztanaga}, {Giannini}, {Gruen}, {Gruendl}, {Harrison}, {Hartley}, {Herner}, {Hoffmann}, {Huang}, {Huff}, {Jain}, {Jarvis}, {Jeffrey}, {Kacprzak}, {Krause}, {Kuropatkin}, {Leget}, {Lemos}, {Liddle}, {MacCrann}, {McCullough}, {Muir}, {Myles}, {Navarro-Alsina}, {Omori}, {Pandey}, {Park}, {Porredon}, {Prat}, {Raveri}, {Refregier}, {Rollins}, {Roodman}, {Rosenfeld}, {Ross}, {Rykoff}, {Samuroff}, {S{\'a}nchez}, {Sanchez}, {Sevilla-Noarbe}, {Sheldon}, {Shin}, {Troja}, {Troxel}, {Tutusaus}, {Varga}, {Wechsler}, {Yanny}, {Yin}, {Zhang}, {Zuntz}, {Aguena}, {Annis}, {Bacon}, {Bertin}, {Bocquet}, {Brooks}, {Burke}, {Carretero}, {Conselice}, {Costanzi}, {da Costa}, {De Vicente}, {Desai}, {Doel}, {Ferrero}, {Flaugher}, {Frieman}, {Garc{\'\i}a-Bellido}, {Gerdes}, {Giannantonio}, {Gschwend}, {Gutierrez}, {Hinton}, {Hollowood}, {Honscheid}, {James}, {Kuehn}, {Lahav}, {March}, {Marshall}, {Melchior}, {Menanteau}, {Miquel}, {Mohr}, {Morgan}, {Paz-Chinch{\'o}n}, {Pieres}, {Sanchez}, {Smith}, {Suchyta}, {Swanson}, {Tarle}, {Thomas}, {To}, \& {DES Collaboration}}]{chen2023constraining}
{Chen}, A., {Aric{\`o}}, G., {Huterer}, D., {et~al.} 2023, \mnras, 518, 5340, \dodoi{10.1093/mnras/stac3213}

\bibitem[{{CHIME/FRB Collaboration} {et~al.}(2020){CHIME/FRB Collaboration}, {Amiri}, {Andersen}, {Band ura}, {Bhardwaj}, {Boyle}, {Brar}, {Chawla}, {Chen}, {Cliche}, {Cubranic}, {Deng}, {Denman}, {Dobbs}, {Dong}, {Fand ino}, {Fonseca}, {Gaensler}, {Giri}, {Good}, {Halpern}, {Hessels}, {Hill}, {H{\"o}fer}, {Josephy}, {Kania}, {Karuppusamy}, {Kaspi}, {Keimpema}, {Kirsten}, {Landecker}, {Lang}, {Leung}, {Li}, {Lin}, {Marcote}, {Masui}, {McKinven}, {Mena-Parra}, {Merryfield}, {Michilli}, {Milutinovic}, {Mirhosseini}, {Naidu}, {Newburgh}, {Ng}, {Nimmo}, {Paragi}, {Patel}, {Pen}, {Pinsonneault-Marotte}, {Pleunis}, {Rafiei-Ravandi}, {Rahman}, {Ransom}, {Renard}, {Sanghavi}, {Scholz}, {Shaw}, {Shin}, {Siegel}, {Singh}, {Smegal}, {Smith}, {Stairs}, {Tendulkar}, {Tretyakov}, {Vanderlinde}, {Wang}, {Wang}, {Wulf}, {Yadav}, \& {Zwaniga}}]{collaboration2020periodic}
{CHIME/FRB Collaboration}, {Amiri}, M., {Andersen}, B.~C., {et~al.} 2020, \nat, 582, 351, \dodoi{10.1038/s41586-020-2398-2}

\bibitem[{{CHIME/FRB Collaboration} {et~al.}(2025{\natexlab{a}}){CHIME/FRB Collaboration}, {Amiri}, {Amouyal}, {Andersen}, {Andrew}, {Bandura}, {Bhardwaj}, {Boyle}, {Brar}, {Cassity}, {Chatterjee}, {Curtin}, {Dobbs}, {Dong}, {Dong}, {Eadie}, {Eftekhari}, {Fong}, {Fonseca}, {Gaensler}, {Halpern}, {Hessels}, {Hopkins}, {Ibik}, {Joseph}, {Kaczmarek}, {Kahinga}, {Kaspi}, {Khairy}, {Kilpatrick}, {Lanman}, {Lazda}, {Leung}, {Main}, {Mas-Ribas}, {Masui}, {Mckinven}, {Mena-Parra}, {Meyers}, {Michilli}, {Milutinovic}, {Nimmo}, {Noble}, {Pandhi}, {Shivraj Patil}, {Pearlman}, {Petroff}, {Pleunis}, {Prochaska}, {Rafiei-Ravandi}, {Rahman}, {Renard}, {Sammons}, {Sand}, {Scholz}, {Shah}, {Shin}, {Siegel}, {Simha}, {Smith}, {Stairs}, {Vanderlinde}, {Wang}, {Wulf}, \& {Zegmott}}]{chime2025catalog}
{CHIME/FRB Collaboration}, {Amiri}, M., {Amouyal}, D., {et~al.} 2025{\natexlab{a}}, arXiv e-prints, arXiv:2502.11217.
\newblock \doarXiv{2502.11217}

\bibitem[{{CHIME/FRB Collaboration} {et~al.}(2025{\natexlab{b}}){CHIME/FRB Collaboration}, {Amiri}, {Andersen}, {Andrew}, {Bandura}, {Bhardwaj}, {Bhopi}, {Bidula}, {Boyle}, {Brar}, {Carlson}, {Cassanelli}, {Cassity}, {Chatterjee}, {Cliche}, {Curtin}, {Darlinger}, {DeBoer}, {Dobbs}, {Dong}, {Eadie}, {Fonseca}, {Gaensler}, {Gusinskaia}, {Halpern}, {Hendricksen}, {Hessels}, {Joseph}, {Kaczmarek}, {Kaspi}, {Khairy}, {Landecker}, {Lanman}, {Kit Lau}, {Lazda}, {Leung}, {Main}, {Masui}, {Mckinven}, {Mena-Parra}, {Meyers}, {Michilli}, {Milutinovic}, {Nimmo}, {Noble}, {Pandhi}, {Pearlman}, {Peterson}, {Petroff}, {Pleunis}, {Pollak}, {Rafiei-Ravandi}, {Renard}, {Sammons}, {Sand}, {Sanghavi}, {Scholz}, {Shah}, {Shin}, {Siegel}, {Siemion}, {Sievers}, {Smith}, {Spear}, {Stairs}, {Vanderlinde}, {Wang}, {Willis}, \& {Zegmott}}]{collaboration2025chimeoutriggers}
{CHIME/FRB Collaboration}, {Amiri}, M., {Andersen}, B.~C., {et~al.} 2025{\natexlab{b}}, arXiv e-prints, arXiv:2504.05192, \dodoi{10.48550/arXiv.2504.05192}

\bibitem[{{Chittidi} {et~al.}(2021){Chittidi}, {Simha}, {Mannings}, {Prochaska}, {Ryder}, {Rafelski}, {Neeleman}, {Macquart}, {Tejos}, {Jorgenson}, {Day}, {Marnoch}, {Bhandari}, {Deller}, {Qiu}, {Bannister}, {Shannon}, \& {Heintz}}]{chittidi2021dissecting}
{Chittidi}, J.~S., {Simha}, S., {Mannings}, A., {et~al.} 2021, \apj, 922, 173, \dodoi{10.3847/1538-4357/ac2818}

\bibitem[{{Connor} \& {Ravi}(2022)}]{connor2022stellar}
{Connor}, L., \& {Ravi}, V. 2022, arXiv e-prints, arXiv:2206.14310.
\newblock \doarXiv{2206.14310}

\bibitem[{{Connor} {et~al.}(2023){Connor}, {Ravi}, {Catha}, {Chen}, {Faber}, {Lamb}, {Hallinan}, {Harnach}, {Hellbourg}, {Hobbs}, {Hodge}, {Hodges}, {Law}, {Rasmussen}, {Sayers}, {Sharma}, {Sherman}, {Shi}, {Simard}, {Somalwar}, {Squillace}, {Weinreb}, {Woody}, \& {Yadlapalli}}]{connor2023deep}
{Connor}, L., {Ravi}, V., {Catha}, M., {et~al.} 2023, arXiv e-prints, arXiv:2302.14788, \dodoi{10.48550/arXiv.2302.14788}

\bibitem[{{Connor} {et~al.}(2024){Connor}, {Ravi}, {Sharma}, {Ocker}, {Faber}, {Hallinan}, {Harnach}, {Hellbourg}, {Hobbs}, {Hodge}, {Hodges}, {Kosogorov}, {Lamb}, {Law}, {Rasmussen}, {Sherman}, {Somalwar}, {Weinreb}, \& {Woody}}]{connor2024gas}
{Connor}, L., {Ravi}, V., {Sharma}, K., {et~al.} 2024, arXiv e-prints, arXiv:2409.16952, \dodoi{10.48550/arXiv.2409.16952}

\bibitem[{{Cook} {et~al.}(2023{\natexlab{a}}){Cook}, {Bhardwaj}, {Gaensler}, {Scholz}, {Eadie}, {Hill}, {Kaspi}, {Masui}, {Curtin}, {Dong}, {Fonseca}, {Herrera-Martin}, {Kaczmarek}, {Lanman}, {Lazda}, {Leung}, {Meyers}, {Michilli}, {Pandhi}, {Pearlman}, {Pleunis}, {Ransom}, {Rahman}, {Sand}, {Shin}, {Smith}, {Stairs}, \& {Stenning}}]{cook2023frb}
{Cook}, A.~M., {Bhardwaj}, M., {Gaensler}, B.~M., {et~al.} 2023{\natexlab{a}}, \apj, 946, 58, \dodoi{10.3847/1538-4357/acbbd0}

\bibitem[{{Cook} {et~al.}(2023{\natexlab{b}}){Cook}, {Mazzarella}, {Helou}, {Alcala}, {Chen}, {Ebert}, {Frayer}, {Kim}, {Lo}, {Madore}, {Ogle}, {Schmitz}, {Singer}, {Terek}, {Valladon}, \& {Wu}}]{cook2023completeness}
{Cook}, D.~O., {Mazzarella}, J.~M., {Helou}, G., {et~al.} 2023{\natexlab{b}}, \apjs, 268, 14, \dodoi{10.3847/1538-4365/acdd06}

\bibitem[{{Cordes} \& {Chatterjee}(2019)}]{cordes2019fast}
{Cordes}, J.~M., \& {Chatterjee}, S. 2019, \araa, 57, 417, \dodoi{10.1146/annurev-astro-091918-104501}

\bibitem[{{Cordes} \& {Lazio}(2002)}]{cordes2002ne2001}
{Cordes}, J.~M., \& {Lazio}, T.~J.~W. 2002, arXiv e-prints, astro.
\newblock \doarXiv{astro-ph/0207156}

\bibitem[{{Cordes} {et~al.}(2016){Cordes}, {Wharton}, {Spitler}, {Chatterjee}, \& {Wasserman}}]{cordes2016radio}
{Cordes}, J.~M., {Wharton}, R.~S., {Spitler}, L.~G., {Chatterjee}, S., \& {Wasserman}, I. 2016, arXiv e-prints, arXiv:1605.05890.
\newblock \doarXiv{1605.05890}

\bibitem[{{Dav{\'e}} {et~al.}(2019){Dav{\'e}}, {Angl{\'e}s-Alc{\'a}zar}, {Narayanan}, {Li}, {Rafieferantsoa}, \& {Appleby}}]{dave2019simba}
{Dav{\'e}}, R., {Angl{\'e}s-Alc{\'a}zar}, D., {Narayanan}, D., {et~al.} 2019, \mnras, 486, 2827, \dodoi{10.1093/mnras/stz937}

\bibitem[{{Dekel} \& {Birnboim}(2006)}]{dekel2006galaxy}
{Dekel}, A., \& {Birnboim}, Y. 2006, \mnras, 368, 2, \dodoi{10.1111/j.1365-2966.2006.10145.x}

\bibitem[{{Driessen} {et~al.}(2024){Driessen}, {Barr}, {Buckley}, {Caleb}, {Chen}, {Chen}, {Gromadzki}, {Jankowski}, {Kraan-Korteweg}, {Palmerio}, {Rajwade}, {Tremou}, {Kramer}, {Stappers}, {Vergani}, {Woudt}, {Bezuidenhout}, {Malenta}, {Morello}, {Sanidas}, {Surnis}, \& {Fender}}]{driessen2024frb}
{Driessen}, L.~N., {Barr}, E.~D., {Buckley}, D.~A.~H., {et~al.} 2024, \mnras, 527, 3659, \dodoi{10.1093/mnras/stad3329}

\bibitem[{{Eftekhari} {et~al.}(2024){Eftekhari}, {Dong}, {Fong}, {Shah}, {Simha}, {Andersen}, {Andrew}, {Bhardwaj}, {Cassanelli}, {Chatterjee}, {Coulter}, {Fonseca}, {Gaensler}, {Gordon}, {Hessels}, {Ibik}, {Joseph}, {Kahinga}, {Kaspi}, {Kharel}, {Kilpatrick}, {Lanman}, {Lazda}, {Leung}, {Liu}, {Mas-Ribas}, {Masui}, {Mckinven}, {Mena-Parra}, {Miller}, {Nimmo}, {Pandhi}, {Pearlman}, {Pleunis}, {Prochaska}, {Rafiei-Ravandi}, {Sammons}, {Scholz}, {Shin}, {Smith}, {Stairs}, \& {Swarali Shivraj}}]{eftekhari2024massive}
{Eftekhari}, T., {Dong}, Y., {Fong}, W., {et~al.} 2024, arXiv e-prints, arXiv:2410.23336, \dodoi{10.48550/arXiv.2410.23336}

\bibitem[{{Euclid Collaboration} {et~al.}(2025){Euclid Collaboration}, {Mellier}, {Abdurro'uf}, {Acevedo Barroso}, {Ach{\'u}carro}, {Adamek}, {Adam}, {Addison}, {Aghanim}, {Aguena}, {Ajani}, {Akrami}, {Al-Bahlawan}, {Alavi}, {Albuquerque}, {Alestas}, {Alguero}, {Allaoui}, {Allen}, {Allevato}, {Alonso-Tetilla}, {Altieri}, {Alvarez-Candal}, {Alvi}, {Amara}, {Amendola}, {Amiaux}, {Andika}, {Andreon}, {Andrews}, {Angora}, {Angulo}, {Annibali}, {Anselmi}, {Anselmi}, {Arcari}, {Archidiacono}, {Aric{\`o}}, {Arnaud}, {Arnouts}, {Asgari}, {Asorey}, {Atayde}, {Atek}, {Atrio-Barandela}, {Aubert}, {Aubourg}, {Auphan}, {Auricchio}, {Aussel}, {Aussel}, {Avelino}, {Avgoustidis}, {Avila}, {Awan}, {Azzollini}, {Baccigalupi}, {Bachelet}, {Bacon}, {Baes}, {Bagley}, {Bahr-Kalus}, {Balaguera-Antolinez}, {Balbinot}, {Balcells}, {Baldi}, {Baldry}, {Balestra}, {Ballardini}, {Ballester}, {Balogh}, {Ba{\~n}ados}, {Barbier}, {Bardelli}, {Baron}, {Barreiro}, {Barrena}, {Barriere}, {Barros}, {Barthelemy}, {Bartolo}, {Basset}, {Battaglia}, {Battisti}, {Baugh}, {Baumont}, {Bazzanini}, {Beaulieu}, {Beckmann}, {Belikov}, {Bel}, {Bellagamba}, {Bella}, {Bellini}, {Benabed}, {Bender}, {Benevento}, {Bennett}, {Benson}, {Bergamini}, {Bermejo-Climent}, {Bernardeau}, {Bertacca}, {Berthe}, {Berthier}, {Bethermin}, {Beutler}, {Bevillon}, {Bhargava}, {Bhatawdekar}, {Bianchi}, {Bisigello}, {Biviano}, {Blake}, {Blanchard}, {Blazek}, {Blot}, {Bosco}, {Bodendorf}, {Boenke}, {B{\"o}hringer}, {Boldrini}, {Bolzonella}, {Bonchi}, {Bonici}, {Bonino}, {Bonino}, {Bonvin}, {Bon}, {Booth}, {Borgani}, {Borlaff}, {Borsato}, {Bose}, {Botticella}, {Boucaud}, {Bouche}, {Boucher}, {Boutigny}, {Bouvard}, {Bouwens}, {Bouy}, {Bowler}, {Bozza}, {Bozzo}, {Branchini}, {Brando}, {Brau-Nogue}, {Brekke}, {Bremer}, {Brescia}, {Breton}, {Brinchmann}, {Brinckmann}, {Brockley-Blatt}, {Brodwin}, {Brouard}, {Brown}, {Bruton}, {Bucko}, {Buddelmeijer}, {Buenadicha}, {Buitrago}, {Burger}, {Burigana}, {Busillo}, {Busonero}, {Cabanac}, {Cabayol-Garcia}, {Cagliari}, {Caillat}, {Caillat}, {Calabrese}, {Calabro}, {Calderone}, {Calura}, {Camacho Quevedo}, {Camera}, {Campos}, {Ca{\~n}as-Herrera}, {Candini}, {Cantiello}, {Capobianco}, {Cappellaro}, {Cappelluti}, {Cappi}, {Caputi}, {Cara}, {Carbone}, {Cardone}, {Carella}, {Carlberg}, {Carle}, {Carminati}, {Caro}, {Carrasco}, {Carretero}, {Carrilho}, {Carron Duque}, \& {Carry}}]{euclid2025overview}
{Euclid Collaboration}, {Mellier}, Y., {Abdurro'uf}, {et~al.} 2025, \aap, 697, A1, \dodoi{10.1051/0004-6361/202450810}

\bibitem[{{Foreman-Mackey} {et~al.}(2013){Foreman-Mackey}, {Hogg}, {Lang}, \& {Goodman}}]{foreman_mackey2013emcee}
{Foreman-Mackey}, D., {Hogg}, D.~W., {Lang}, D., \& {Goodman}, J. 2013, \pasp, 125, 306, \dodoi{10.1086/670067}

\bibitem[{{Giodini} {et~al.}(2013){Giodini}, {Lovisari}, {Pointecouteau}, {Ettori}, {Reiprich}, \& {Hoekstra}}]{giodini2013}
{Giodini}, S., {Lovisari}, L., {Pointecouteau}, E., {et~al.} 2013, \ssr, 177, 247, \dodoi{10.1007/s11214-013-9994-5}

\bibitem[{{Glowacki} {et~al.}(2023){Glowacki}, {Lee-Waddell}, {Deller}, {Deg}, {Gordon}, {Grundy}, {Marnoch}, {Shen}, {Ryder}, {Shannon}, {Wong}, {D{\'e}nes}, {Koribalski}, {Murugeshan}, {Rhee}, {Westmeier}, {Bhandari}, {Bosma}, {Holwerda}, \& {Prochaska}}]{glowacki2023wallaby}
{Glowacki}, M., {Lee-Waddell}, K., {Deller}, A.~T., {et~al.} 2023, \apj, 949, 25, \dodoi{10.3847/1538-4357/acc1e3}

\bibitem[{{Glowacki} {et~al.}(2025){Glowacki}, {Bera}, {James}, {Paterson}, {Deller}, {Gordon}, {Marnoch}, {Muller}, {Prochaska}, {Ryder}, {Shannon}, {Tejos}, \& {Mannings}}]{glowacki2025investigation}
{Glowacki}, M., {Bera}, A., {James}, C.~W., {et~al.} 2025, arXiv e-prints, arXiv:2506.23403, \dodoi{10.48550/arXiv.2506.23403}

\bibitem[{{Gordon} {et~al.}(2023){Gordon}, {Fong}, {Kilpatrick}, {Eftekhari}, {Leja}, {Prochaska}, {Nugent}, {Bhandari}, {Blanchard}, {Caleb}, {Day}, {Deller}, {Dong}, {Glowacki}, {Gourdji}, {Mannings}, {Mahoney}, {Marnoch}, {Miller}, {Paterson}, {Rastinejad}, {Ryder}, {Sadler}, {Scott}, {Sears}, {Shannon}, {Simha}, {Stappers}, \& {Tejos}}]{gordon2023demographics}
{Gordon}, A.~C., {Fong}, W.-f., {Kilpatrick}, C.~D., {et~al.} 2023, \apj, 954, 80, \dodoi{10.3847/1538-4357/ace5aa}

\bibitem[{{Gordon} {et~al.}(2025){Gordon}, {Fong}, {Deller}, {Marnoch}, {Lim}, {Peng}, {Bannister}, {Bera}, {Bhat}, {Dial}, {Dong}, {Eftekhari}, {Glowacki}, {Gourdji}, {Gupta}, {Jahns-Schindler}, {Jaini}, {Kilpatrick}, {Liu}, {Prochaska}, {Ryder}, {Shannon}, {Simha}, {Tejos}, {Wang}, \& {Wang}}]{gordon2025mapping}
{Gordon}, A.~C., {Fong}, W.-f., {Deller}, A.~T., {et~al.} 2025, arXiv e-prints, arXiv:2506.06453, \dodoi{10.48550/arXiv.2506.06453}

\bibitem[{{Gordon} {et~al.}(2016){Gordon}, {Brisken}, \& {Max-Moerbeck}}]{gordon2016difxcalc}
{Gordon}, D., {Brisken}, W., \& {Max-Moerbeck}, W. 2016, in New Horizons with VGOS, ed. D.~{Behrend}, K.~D. {Baver}, \& K.~L. {Armstrong}, 187--192

\bibitem[{{Hadzhiyska} {et~al.}(2024){Hadzhiyska}, {Ferraro}, {Ried Guachalla}, {Schaan}, {Aguilar}, {Battaglia}, {Bond}, {Brooks}, {Calabrese}, {Choi}, {Claybaugh}, {Coulton}, {Dawson}, {Devlin}, {Dey}, {Doel}, {Duivenvoorden}, {Dunkley}, {Farren}, {Font-Ribera}, {Forero-Romero}, {Gallardo}, {Gazta{\~n}aga}, {Gontcho Gontcho}, {Gralla}, {Le Guillou}, {Gutierrez}, {Guy}, {Hill}, {Hlo{\v{z}}ek}, {Honscheid}, {Juneau}, {Kisner}, {Kremin}, {Landriau}, {Liu}, {Louis}, {MacCrann}, {de Macorra}, {Madhavacheril}, {Manera}, {Meisner}, {Miquel}, {Moodley}, {Moustakas}, {Mroczkowski}, {Naess}, {Newman}, {Niemack}, {Niz}, {Page}, {Palanque-Delabrouille}, {Partridge}, {Percival}, {Prada}, {Qu}, {Rossi}, {Sanchez}, {Schlegel}, {Schubnell}, {Sehgal}, {Seo}, {Sif{\'o}n}, {Spergel}, {Sprayberry}, {Staggs}, {Tarl{\'e}}, {Vargas}, {Vavagiakis}, {Weaver}, {Wollack}, {Zhou}, \& {Zou}}]{hadzhiyska2024evidence}
{Hadzhiyska}, B., {Ferraro}, S., {Ried Guachalla}, B., {et~al.} 2024, arXiv e-prints, arXiv:2407.07152, \dodoi{10.48550/arXiv.2407.07152}

\bibitem[{{Hallinan} {et~al.}(2019){Hallinan}, {Ravi}, {Weinreb}, {Kocz}, {Huang}, {Woody}, {Lamb}, {D'Addario}, {Catha}, {Law}, {Kulkarni}, {Phinney}, {Eastwood}, {Bouman}, {McLaughlin}, {Ransom}, {Siemens}, {Cordes}, {Lynch}, {Kaplan}, {Brazier}, {Bhatnagar}, {Myers}, {Walter}, \& {Gaensler}}]{hallinan2019dsa}
{Hallinan}, G., {Ravi}, V., {Weinreb}, S., {et~al.} 2019, in Bulletin of the American Astronomical Society, Vol.~51, 255, \dodoi{10.48550/arXiv.1907.07648}

\bibitem[{Harris {et~al.}(2020)Harris, Millman, van~der Walt, Gommers, Virtanen, Cournapeau, Wieser, Taylor, Berg, Smith, Kern, Picus, Hoyer, van Kerkwijk, Brett, Haldane, del R{\'{i}}o, Wiebe, Peterson, G{\'{e}}rard-Marchant, Sheppard, Reddy, Weckesser, Abbasi, Gohlke, \& Oliphant}]{harris2020array}
Harris, C.~R., Millman, K.~J., van~der Walt, S.~J., {et~al.} 2020, Nature, 585, 357, \dodoi{10.1038/s41586-020-2649-2}

\bibitem[{{Hewitt} {et~al.}(2024){Hewitt}, {Bhardwaj}, {Gordon}, {Kirichenko}, {Nimmo}, {Bhandari}, {Cognard}, {Fong}, {Gil de Paz}, {Gopinath}, {Hessels}, {Kirsten}, {Marcote}, {Bezrukovs}, {Blaauw}, {Bray}, {Buttaccio}, {Cassanelli}, {Chawla}, {Corongiu}, {Deng}, {Didehbani}, {Dong}, {Gawro{\'n}ski}, {Giroletti}, {Guillemot}, {Huang}, {Ivanov}, {Joseph}, {Kaspi}, {Kharinov}, {Lazda}, {Lindqvist}, {Maccaferri}, {Mas-Ribas}, {Masui}, {Mckinven}, {Melnikov}, {Michilli}, {Mikhailov}, {Nugent}, {Ould-Boukattine}, {Paragi}, {Pearlman}, {Pen}, {Pleunis}, {Sand}, {Shah}, {Shin}, {Snelders}, {Venturi}, {Wang}, {Williams-Baldwin}, {Yang}, \& {Yuan}}]{hewitt2024repeating}
{Hewitt}, D.~M., {Bhardwaj}, M., {Gordon}, A.~C., {et~al.} 2024, \apjl, 977, L4, \dodoi{10.3847/2041-8213/ad8ce1}

\bibitem[{{Hsu} {et~al.}(2025){Hsu}, {Hashimoto}, {Yang}, {Yamasaki}, {Goto}, {Lo}, {Wang}, {Lin}, {Ho}, \& {Raquel}}]{hsu2025decoding}
{Hsu}, T.-Y., {Hashimoto}, T., {Yang}, T.-C., {et~al.} 2025, \aap, 698, A163, \dodoi{10.1051/0004-6361/202452714}

\bibitem[{{Huang} {et~al.}(2012){Huang}, {Haynes}, {Giovanelli}, \& {Brinchmann}}]{huang2012arecibo}
{Huang}, S., {Haynes}, M.~P., {Giovanelli}, R., \& {Brinchmann}, J. 2012, \apj, 756, 113, \dodoi{10.1088/0004-637X/756/2/113}

\bibitem[{{Huang} {et~al.}(2025){Huang}, {Simha}, {Khrykin}, {Lee}, {Prochaska}, {Tejos}, {Bannister}, {Barrios}, {Chisholm}, {Cooke}, {Deller}, {Glowacki}, {Marnoch}, {Shannon}, \& {Zhang}}]{huang2025frb}
{Huang}, Y., {Simha}, S., {Khrykin}, I.~S., {et~al.} 2025, \apjs, 277, 64, \dodoi{10.3847/1538-4365/adbc7f}

\bibitem[{Hunter(2007)}]{hunter2007matplotlib}
Hunter, J.~D. 2007, Computing in Science \& Engineering, 9, 90, \dodoi{10.1109/MCSE.2007.55}

\bibitem[{{Hussaini} {et~al.}(2025){Hussaini}, {Connor}, {Konietzka}, {Ravi}, {Faber}, {Sharma}, \& {Sherman}}]{hussaini2025correlation}
{Hussaini}, M., {Connor}, L., {Konietzka}, R.~M., {et~al.} 2025, arXiv e-prints, arXiv:2506.04186, \dodoi{10.48550/arXiv.2506.04186}

\bibitem[{{Ibik} {et~al.}(2024a){Ibik}, {Drout}, {Gaensler}, {Scholz}, {Michilli}, {Bhardwaj}, {Kaspi}, {Pleunis}, {Cassanelli}, {Cook}, {Dong}, {Kaczmarek}, {Leung}, {Lu}, {Masui}, {Pearlman}, {Rafiei-Ravandi}, {Sand}, {Shin}, {Smith}, \& {Stairs}}]{Ibik2024host}
{Ibik}, A.~L., {Drout}, M.~R., {Gaensler}, B.~M., {et~al.} 2024a, \apj, 961, 99, \dodoi{10.3847/1538-4357/ad0893}

\bibitem[{{Jahns-Schindler} {et~al.}(2023){Jahns-Schindler}, {Spitler}, {Walker}, \& {Baugh}}]{jahnsschindler2023how}
{Jahns-Schindler}, J.~N., {Spitler}, L.~G., {Walker}, C. R.~H., \& {Baugh}, C.~M. 2023, \mnras, 523, 5006, \dodoi{10.1093/mnras/stad1659}

\bibitem[{James {et~al.}(2021)James, Prochaska, Macquart, North-Hickey, Bannister, \& Dunning}]{james2021z}
James, C.~W., Prochaska, J.~X., Macquart, J.-P., {et~al.} 2021, Monthly Notices of the Royal Astronomical Society, 509, 4775, \dodoi{10.1093/mnras/stab3051}

\bibitem[{{James} {et~al.}(2022){James}, {Prochaska}, {Macquart}, {North-Hickey}, {Bannister}, \& {Dunning}}]{james2022fast}
{James}, C.~W., {Prochaska}, J.~X., {Macquart}, J.~P., {et~al.} 2022, \mnras, 510, L18, \dodoi{10.1093/mnrasl/slab117}

\bibitem[{{Jaroszynski}(2019)}]{jaroszynski2019fast}
{Jaroszynski}, M. 2019, \mnras, 484, 1637, \dodoi{10.1093/mnras/sty3529}

\bibitem[{{Keating} \& {Pen}(2020)}]{keating2020exploring}
{Keating}, L.~C., \& {Pen}, U.-L. 2020, \mnras, 496, L106, \dodoi{10.1093/mnrasl/slaa095}

\bibitem[{{Khrykin} {et~al.}(2024{\natexlab{a}}){Khrykin}, {Sorini}, {Lee}, \& {Dav{\'e}}}]{khrykin2024cosmic}
{Khrykin}, I.~S., {Sorini}, D., {Lee}, K.-G., \& {Dav{\'e}}, R. 2024{\natexlab{a}}, \mnras, 529, 537, \dodoi{10.1093/mnras/stae525}

\bibitem[{{Khrykin} {et~al.}(2024{\natexlab{b}}){Khrykin}, {Ata}, {Lee}, {Simha}, {Huang}, {Prochaska}, {Tejos}, {Bannister}, {Cooke}, {Day}, {Deller}, {Glowacki}, {Gordon}, {James}, {Marnoch}, {Shannon}, {Zhang}, \& {Bernales-Cortes}}]{khrykin2024flimflam}
{Khrykin}, I.~S., {Ata}, M., {Lee}, K.-G., {et~al.} 2024{\natexlab{b}}, \apj, 973, 151, \dodoi{10.3847/1538-4357/ad6567}

\bibitem[{{Kirsten} {et~al.}(2022){Kirsten}, {Marcote}, {Nimmo}, {Hessels}, {Bhardwaj}, {Tendulkar}, {Keimpema}, {Yang}, {Snelders}, {Scholz}, {Pearlman}, {Law}, {Peters}, {Giroletti}, {Paragi}, {Bassa}, {Hewitt}, {Bach}, {Bezrukovs}, {Burgay}, {Buttaccio}, {Conway}, {Corongiu}, {Feiler}, {Forss{\'e}n}, {Gawro{\'n}ski}, {Karuppusamy}, {Kharinov}, {Lindqvist}, {Maccaferri}, {Melnikov}, {Ould-Boukattine}, {Possenti}, {Surcis}, {Wang}, {Yuan}, {Aggarwal}, {Anna-Thomas}, {Bower}, {Blaauw}, {Burke-Spolaor}, {Cassanelli}, {Clarke}, {Fonseca}, {Gaensler}, {Gopinath}, {Kaspi}, {Kassim}, {Lazio}, {Leung}, {et~al.}}]{kirsten2022repeating}
{Kirsten}, F., {Marcote}, B., {Nimmo}, K., {et~al.} 2022, \nat, 602, 585, \dodoi{10.1038/s41586-021-04354-w}

\bibitem[{{Kourkchi} {et~al.}(2020){Kourkchi}, {Tully}, {Eftekharzadeh}, {Llop}, {Courtois}, {Guinet}, {Dupuy}, {Neill}, {Seibert}, {Andrews}, {Chuang}, {Danesh}, {Gonzalez}, {Holthaus}, {Mokelke}, {Schoen}, \& {Urasaki}}]{kourkchi2020cosmicflows}
{Kourkchi}, E., {Tully}, R.~B., {Eftekharzadeh}, S., {et~al.} 2020, \apj, 902, 145, \dodoi{10.3847/1538-4357/abburn1966depolarizationb}

\bibitem[{{Kravtsov}(2013)}]{kravtsov2013size}
{Kravtsov}, A.~V. 2013, \apjl, 764, L31, \dodoi{10.1088/2041-8205/764/2/L31}

\bibitem[{{Lanman} {et~al.}(2022){Lanman}, {Andersen}, {Chawla}, {Josephy}, {Noble}, {Kaspi}, {Bandura}, {Bhardwaj}, {Boyle}, {Brar}, {Breitman}, {Cassanelli}, {Dong}, {Fonseca}, {Gaensler}, {Good}, {Kaczmarek}, {Leung}, {Masui}, {Meyers}, {Ng}, {Patel}, {Pearlman}, {Petroff}, {Pleunis}, {Rafiei-Ravandi}, {Rahman}, {Sanghavi}, {Scholz}, {Shin}, {Stairs}, {Tendulkar}, \& {Zwaniga}}]{lanman2022sudden}
{Lanman}, A.~E., {Andersen}, B.~C., {Chawla}, P., {et~al.} 2022, \apj, 927, 59, \dodoi{10.3847/1538-4357/ac4bc7}

\bibitem[{{Law} {et~al.}(2024){Law}, {Sharma}, {Ravi}, {Chen}, {Catha}, {Connor}, {Faber}, {Hallinan}, {Harnach}, {Hellbourg}, {Hobbs}, {Hodge}, {Hodges}, {Lamb}, {Rasmussen}, {Sherman}, {Shi}, {Simard}, {Squillace}, {Weinreb}, {Woody}, \& {Yurk}}]{law2024deep}
{Law}, C.~J., {Sharma}, K., {Ravi}, V., {et~al.} 2024, \apj, 967, 29, \dodoi{10.3847/1538-4357/ad3736}

\bibitem[{{Lee} {et~al.}(2022){Lee}, {Ata}, {Khrykin}, {Huang}, {Prochaska}, {Cooke}, {Zhang}, \& {Batten}}]{lee2022constraining}
{Lee}, K.-G., {Ata}, M., {Khrykin}, I.~S., {et~al.} 2022, \apj, 928, 9, \dodoi{10.3847/1538-4357/ac4f62}

\bibitem[{{Lee} {et~al.}(2023){Lee}, {Khrykin}, {Simha}, {Ata}, {Huang}, {Prochaska}, {Tejos}, {Cooke}, {Nagamine}, \& {Zhang}}]{lee2023frb}
{Lee}, K.-G., {Khrykin}, I.~S., {Simha}, S., {et~al.} 2023, \apjl, 954, L7, \dodoi{10.3847/2041-8213/acefb5}

\bibitem[{{Lee-Waddell} {et~al.}(2023){Lee-Waddell}, {James}, {Ryder}, {Mahony}, {Bahramian}, {Koribalski}, {Kumar}, {Marnoch}, {North-Hickey}, {Sadler}, {Shannon}, {Tejos}, {Thorne}, {Wang}, \& {Wayth}}]{leewadell2023host}
{Lee-Waddell}, K., {James}, C.~W., {Ryder}, S.~D., {et~al.} 2023, \pasa, 40, e029, \dodoi{10.1017/pasa.2023.27}

\bibitem[{{Leung} {et~al.}(2024){Leung}, {Andrew}, {Masui}, {Brar}, {Cassanelli}, {Chatterjee}, {Kaspi}, {Khairy}, {Lanman}, {Lazda}, {Mena-Parra}, {Noble}, {Pearlman}, {Rahman}, {Sanghavi}, \& {Shah}}]{leung2024vlbi}
{Leung}, C., {Andrew}, S., {Masui}, K.~W., {et~al.} 2024, arXiv e-prints, arXiv:2403.05631, \dodoi{10.48550/arXiv.2403.05631}

\bibitem[{{Lin} {et~al.}(2022){Lin}, {Lin}, {Li}, {Tseng}, {Jiang}, {Wang}, {Cheng}, {Pen}, {Chen}, {Chen}, {Chen}, {Goto}, {Hashimoto}, {Hwang}, {King}, {Kubo}, {Kuo}, {Mills}, {Nam}, {Oshiro}, {Shen}, {Tseng}, {Wang}, {Wu}, {Bower}, {Chang}, {Chen}, {Chen}, {Chiang}, {Fedynitch}, {Gusinskaia}, {Ho}, {Hsiao}, {Hu}, {Huang}, {J{\'a}uregui Garc{\'\i}a}, {Kim}, {Kuo}, {Ling}, {On}, {Peterson}, {R. Raquel}, {Su}, {Uno}, {Wu}, {Yamasaki}, \& {Zhu}}]{lin2022burstt}
{Lin}, H.-H., {Lin}, K.-y., {Li}, C.-T., {et~al.} 2022, \pasp, 134, 094106, \dodoi{10.1088/1538-3873/ac8f71}

\bibitem[{{Lorimer} \& {Kramer}(2004)}]{lorimer2004handbook}
{Lorimer}, D.~R., \& {Kramer}, M. 2004, {Handbook of Pulsar Astronomy}, Vol.~4 (Cambridge University Press)

\bibitem[{{LSST Science Collaboration} {et~al.}(2009){LSST Science Collaboration}, {Abell}, {Allison}, {Anderson}, {Andrew}, {Angel}, {Armus}, {Arnett}, {Asztalos}, {Axelrod}, {Bailey}, {Ballantyne}, {Bankert}, {Barkhouse}, {Barr}, {Barrientos}, {Barth}, {Bartlett}, {Becker}, {Becla}, {Beers}, {Bernstein}, {Biswas}, {Blanton}, {Bloom}, {Bochanski}, {Boeshaar}, {Borne}, {Bradac}, {Brandt}, {Bridge}, {Brown}, {Brunner}, {Bullock}, {Burgasser}, {Burge}, {Burke}, {Cargile}, {Chandrasekharan}, {Chartas}, {Chesley}, {Chu}, {Cinabro}, {Claire}, {Claver}, {Clowe}, {Connolly}, {Cook}, {Cooke}, {Cooray}, {Covey}, {Culliton}, {de Jong}, {de Vries}, {Debattista}, {Delgado}, {Dell'Antonio}, {Dhital}, {Di Stefano}, {Dickinson}, {Dilday}, {Djorgovski}, {Dobler}, {Donalek}, {Dubois-Felsmann}, {Durech}, {Eliasdottir}, {Eracleous}, {Eyer}, {Falco}, {Fan}, {Fassnacht}, {Ferguson}, {Fernandez}, {Fields}, {Finkbeiner}, {Figueroa}, {Fox}, {Francke}, {Frank}, {Frieman}, {Fromenteau}, {Furqan}, {Galaz}, {Gal-Yam}, {Garnavich}, {Gawiser}, {Geary}, {Gee}, {Gibson}, {Gilmore}, {Grace}, {Green}, {Gressler}, {Grillmair}, {Habib}, {Haggerty}, {Hamuy}, {Harris}, {Hawley}, {Heavens}, {Hebb}, {Henry}, {Hileman}, {Hilton}, {Hoadley}, {Holberg}, {Holman}, {Howell}, {Infante}, {Ivezic}, {Jacoby}, {Jain}, {R}, {Jedicke}, {Jee}, {Garrett Jernigan}, {Jha}, {Johnston}, {Jones}, {Juric}, {Kaasalainen}, {Styliani}, {Kafka}, {Kahn}, {Kaib}, {Kalirai}, {Kantor}, {Kasliwal}, {Keeton}, {Kessler}, {Knezevic}, {Kowalski}, {Krabbendam}, {Krughoff}, {Kulkarni}, {Kuhlman}, {Lacy}, {Lepine}, {Liang}, {Lien}, {Lira}, {Long}, {Lorenz}, {Lotz}, {Lupton}, {Lutz}, {Macri}, {Mahabal}, {Mandelbaum}, {Marshall}, {May}, {McGehee}, {Meadows}, {Meert}, {Milani}, {Miller}, {Miller}, {Mills}, {Minniti}, {Monet}, {Mukadam}, {Nakar}, {Neill}, {Newman}, {Nikolaev}, {Nordby}, {O'Connor}, {Oguri}, {Oliver}, {Olivier}, {Olsen}, {Olsen}, {Olszewski}, {Oluseyi}, {Padilla}, {Parker}, {Pepper}, {Peterson}, {Petry}, {Pinto}, {Pizagno}, {Popescu}, {Prsa}, {Radcka}, {Raddick}, {Rasmussen}, {Rau}, {Rho}, {Rhoads}, {Richards}, {Ridgway}, {Robertson}, {Roskar}, {Saha}, {Sarajedini}, {Scannapieco}, {Schalk}, {Schindler}, \& {Schmidt}}]{lsst2009science}
{LSST Science Collaboration}, {Abell}, P.~A., {Allison}, J., {et~al.} 2009, arXiv e-prints, arXiv:0912.0201, \dodoi{10.48550/arXiv.0912.0201}

\bibitem[{{Macquart} {et~al.}(2020){Macquart}, {Prochaska}, {McQuinn}, {Bannister}, {Bhandari}, {Day}, {Deller}, {Ekers}, {James}, {Marnoch}, {Os{\l}owski}, {Phillips}, {Ryder}, {Scott}, {Shannon}, \& {Tejos}}]{macquart2020census}
{Macquart}, J.~P., {Prochaska}, J.~X., {McQuinn}, M., {et~al.} 2020, \nat, 581, 391, \dodoi{10.1038/s41586-020-2300-2}

\bibitem[{{Madau} \& {Dickinson}(2014)}]{madau2014cosmic}
{Madau}, P., \& {Dickinson}, M. 2014, \araa, 52, 415, \dodoi{10.1146/annurev-astro-081811-125615}

\bibitem[{{Maddox} {et~al.}(2015){Maddox}, {Hess}, {Obreschkow}, {Jarvis}, \& {Blyth}}]{maddox2015variation}
{Maddox}, N., {Hess}, K.~M., {Obreschkow}, D., {Jarvis}, M.~J., \& {Blyth}, S.~L. 2015, \mnras, 447, 1610, \dodoi{10.1093/mnras/stu2532}

\bibitem[{Madhavacheril {et~al.}(2019)Madhavacheril, Battaglia, Smith, \& Sievers}]{madhavacheril2019cosmology}
Madhavacheril, M.~S., Battaglia, N., Smith, K.~M., \& Sievers, J.~L. 2019, Phys. Rev. D, 100, 103532, \dodoi{10.1103/PhysRevD.100.103532}

\bibitem[{{Mahajan} {et~al.}(2018){Mahajan}, {Drinkwater}, {Driver}, {Hopkins}, {Graham}, {Brough}, {Brown}, {Holwerda}, {Owers}, \& {Pimbblet}}]{mahajan2018galaxy}
{Mahajan}, S., {Drinkwater}, M.~J., {Driver}, S., {et~al.} 2018, \mnras, 475, 788, \dodoi{10.1093/mnras/stx3202}

\bibitem[{{Mannings} {et~al.}(2023){Mannings}, {Pakmor}, {Prochaska}, {van de Voort}, {Simha}, {Shannon}, {Tejos}, {Deller}, \& {Rafelski}}]{mannings2023}
{Mannings}, A.~G., {Pakmor}, R., {Prochaska}, J.~X., {et~al.} 2023, \apj, 954, 179, \dodoi{10.3847/1538-4357/ace7bb}

\bibitem[{{Marcote} {et~al.}(2020){Marcote}, {Nimmo}, {Hessels}, {Tendulkar}, {Bassa}, {Paragi}, {Keimpema}, {Bhardwaj}, {Karuppusamy}, {Kaspi}, {Law}, {Michilli}, {Aggarwal}, {Andersen}, {Archibald}, {Bandura}, {Bower}, {Boyle}, {Brar}, {Burke-Spolaor}, {Butler}, {Cassanelli}, {Chawla}, {Demorest}, {Dobbs}, {Fonseca}, {Giri}, {Good}, {Gourdji}, {Josephy}, {Kirichenko}, {Kirsten}, {Landecker}, {Lang}, {Lazio}, {Li}, {Lin}, {Linford}, {Masui}, {Mena-Parra}, {Naidu}, {Ng}, {Patel}, {Pen}, {Pleunis}, {Rafiei-Ravandi}, {Rahman}, {Renard}, {Scholz}, {Siegel}, {et~al.}}]{marcote2020repeating}
{Marcote}, B., {Nimmo}, K., {Hessels}, J.~W.~T., {et~al.} 2020, \nat, 577, 190, \dodoi{10.1038/s41586-019-1866-z}

\bibitem[{{Marnoch} {et~al.}(2023){Marnoch}, {Ryder}, {James}, {Gordon}, {Sammons}, {Prochaska}, {Tejos}, {Deller}, {Scott}, {Bhandari}, {Glowacki}, {Mahony}, {McDermid}, {Sadler}, {Shannon}, \& {Qiu}}]{marnoch2023unseen}
{Marnoch}, L., {Ryder}, S.~D., {James}, C.~W., {et~al.} 2023, \mnras, 525, 994, \dodoi{10.1093/mnras/stad2353}

\bibitem[{{Masui} \& {Sigurdson}(2015)}]{masui2015dispersion}
{Masui}, K.~W., \& {Sigurdson}, K. 2015, \prl, 115, 121301, \dodoi{10.1103/PhysRevLett.115.121301}

\bibitem[{{McQuinn}(2014)}]{mcquinn2014locating}
{McQuinn}, M. 2014, \apjl, 780, L33, \dodoi{10.1088/2041-8205/780/2/L33}

\bibitem[{{Medlock} {et~al.}(2025{\natexlab{a}}){Medlock}, {Nagai}, {Angl{\'e}s-Alc{\'a}zar}, \& {Gebhardt}}]{medlock2025constraining}
{Medlock}, I., {Nagai}, D., {Angl{\'e}s-Alc{\'a}zar}, D., \& {Gebhardt}, M. 2025{\natexlab{a}}, \apj, 983, 46, \dodoi{10.3847/1538-4357/adbc9c}

\bibitem[{{Medlock} {et~al.}(2024){Medlock}, {Nagai}, {Singh}, {Oppenheimer}, {Angl{\'e}s-Alc{\'a}zar}, \& {Villaescusa-Navarro}}]{medlock2024probing}
{Medlock}, I., {Nagai}, D., {Singh}, P., {et~al.} 2024, \apj, 967, 32, \dodoi{10.3847/1538-4357/ad3070}

\bibitem[{{Medlock} {et~al.}(2025{\natexlab{b}}){Medlock}, {Neufeld}, {Nagai}, {Angl{\'e}s-Alc{\'a}zar}, {Genel}, {Oppenheimer}, {Sims}, {Singh}, \& {Villaescusa-Navarro}}]{2025ApJ...980...61M}
{Medlock}, I., {Neufeld}, C., {Nagai}, D., {et~al.} 2025{\natexlab{b}}, \apj, 980, 61, \dodoi{10.3847/1538-4357/ada442}

\bibitem[{{Michilli} {et~al.}(2018){Michilli}, {Seymour}, {Hessels}, {Spitler}, {Gajjar}, {Archibald}, {Bower}, {Chatterjee}, {Cordes}, {Gourdji}, {Heald}, {Kaspi}, {Law}, {Sobey}, {Adams}, {Bassa}, {Bogdanov}, {Brinkman}, {Demorest}, {Fernandez}, {Hellbourg}, {Lazio}, {Lynch}, {Maddox}, {Marcote}, {McLaughlin}, {Paragi}, {Ransom}, {Scholz}, {Siemion}, {Tendulkar}, {van Rooy}, {Wharton}, \& {Whitlow}}]{michilli2018extreme}
{Michilli}, D., {Seymour}, A., {Hessels}, J.~W.~T., {et~al.} 2018, \nat, 553, 182, \dodoi{10.1038/nature25149}

\bibitem[{{Michilli} {et~al.}(2023){Michilli}, {Bhardwaj}, {Brar}, {Gaensler}, {Kaspi}, {Kirichenko}, {Masui}, {Mckinven}, {Ng}, {Patel}, {Sand}, {Scholz}, {Shin}, {Siegel}, {Stairs}, {Cassanelli}, {Cook}, {Dobbs}, {Dong}, {Fonseca}, {Ibik}, {Kaczmarek}, {Leung}, {Pearlman}, {Petroff}, {Pleunis}, {Rafiei-Ravandi}, {Sanghavi}, {Shaw}, \& {Tendulkar}}]{michilli2023subarcminute}
{Michilli}, D., {Bhardwaj}, M., {Brar}, C., {et~al.} 2023, \apj, 950, 134, \dodoi{10.3847/1538-4357/accf89}

\bibitem[{{Newton} \& {Kay}(2013)}]{newton2013study}
{Newton}, R. D.~A., \& {Kay}, S.~T. 2013, \mnras, 434, 3606, \dodoi{10.1093/mnras/stt1285}

\bibitem[{{Ni} {et~al.}(2023{\natexlab{a}}){Ni}, {Genel}, {Angl{\'e}s-Alc{\'a}zar}, {Villaescusa-Navarro}, {Jo}, {Bird}, {Di Matteo}, {Croft}, {Chen}, {de Santi}, {Gebhardt}, {Shao}, {Pandey}, {Hernquist}, \& {Dave}}]{camels_data_release2}
{Ni}, Y., {Genel}, S., {Angl{\'e}s-Alc{\'a}zar}, D., {et~al.} 2023{\natexlab{a}}, \apj, 959, 136, \dodoi{10.3847/1538-4357/ad022a}

\bibitem[{{Ni} {et~al.}(2023{\natexlab{b}}){Ni}, {Genel}, {Angl{\'e}s-Alc{\'a}zar}, {Villaescusa-Navarro}, {Jo}, {Bird}, {Di Matteo}, {Croft}, {Chen}, {de Santi}, {Gebhardt}, {Shao}, {Pandey}, {Hernquist}, \& {Dave}}]{ni2023camels}
---. 2023{\natexlab{b}}, \apj, 959, 136, \dodoi{10.3847/1538-4357/ad022a}

\bibitem[{{Nicola} {et~al.}(2022){Nicola}, {Villaescusa-Navarro}, {Spergel}, {Dunkley}, {Angl{\'e}s-Alc{\'a}zar}, {Dav{\'e}}, {Genel}, {Hernquist}, {Nagai}, {Somerville}, \& {Wandelt}}]{nicola2022breaking}
{Nicola}, A., {Villaescusa-Navarro}, F., {Spergel}, D.~N., {et~al.} 2022, \jcap, 2022, 046, \dodoi{10.1088/1475-7516/2022/04/046}

\bibitem[{{Niu} {et~al.}(2021){Niu}, {Aggarwal}, {Li}, {Zhang}, {Chatterjee}, {Tsai}, {Yu}, {Law}, {Burke-Spolaor}, {Cordes}, {Zhang}, {Ocker}, {Yao}, {Wang}, {Feng}, {Niino}, {Bochenek}, {Cruces}, {Connor}, {Jiang}, {Dai}, {Luo}, {Li}, {Miao}, {Niu}, {Anna-Thomas}, {Sydnor}, {Stern}, {Wang}, {Yuan}, {Yue}, {Zhou}, {Yan}, {Zhu}, \& {Zhang}}]{niu2021repeating}
{Niu}, C.~H., {Aggarwal}, K., {Li}, D., {et~al.} 2021, arXiv e-prints, arXiv:2110.07418.
\newblock \doarXiv{2110.07418}

\bibitem[{{Ocker} {et~al.}(2020){Ocker}, {Cordes}, \& {Chatterjee}}]{ocker2020electron}
{Ocker}, S.~K., {Cordes}, J.~M., \& {Chatterjee}, S. 2020, \apj, 897, 124, \dodoi{10.3847/1538-4357/ab98f9}

\bibitem[{{Ocker} {et~al.}(2022){Ocker}, {Cordes}, {Chatterjee}, {Niu}, {Li}, {McKee}, {Law}, {Tsai}, {Anna-Thomas}, {Yao}, \& {Cruces}}]{ocker2022large}
{Ocker}, S.~K., {Cordes}, J.~M., {Chatterjee}, S., {et~al.} 2022, \apj, 931, 87, \dodoi{10.3847/1538-4357/ac6504}

\bibitem[{{Orr} {et~al.}(2024){Orr}, {Burkhart}, {Lu}, {Ponnada}, \& {Hummels}}]{orr2024objects}
{Orr}, M.~E., {Burkhart}, B., {Lu}, W., {Ponnada}, S.~B., \& {Hummels}, C.~B. 2024, \apjl, 972, L26, \dodoi{10.3847/2041-8213/ad725b}

\bibitem[{{Parkash} {et~al.}(2018){Parkash}, {Brown}, {Jarrett}, \& {Bonne}}]{parkash2018relationships}
{Parkash}, V., {Brown}, M. J.~I., {Jarrett}, T.~H., \& {Bonne}, N.~J. 2018, \apj, 864, 40, \dodoi{10.3847/1538-4357/aad3b9}

\bibitem[{{Petroff} {et~al.}(2019){Petroff}, {Hessels}, \& {Lorimer}}]{petroff2019fast}
{Petroff}, E., {Hessels}, J.~W.~T., \& {Lorimer}, D.~R. 2019, \aapr, 27, 4, \dodoi{10.1007/s00159-019-0116-6}

\bibitem[{{Petroff} {et~al.}(2022){Petroff}, {Hessels}, \& {Lorimer}}]{petroff2022fast}
---. 2022, \aapr, 30, 2, \dodoi{10.1007/s00159-022-00139-w}

\bibitem[{{Planck Collaboration} {et~al.}(2020){Planck Collaboration}, {Aghanim}, {Akrami}, {Ashdown}, {Aumont}, {Baccigalupi}, {Ballardini}, {Banday}, {Barreiro}, {Bartolo}, {Basak}, {Benabed}, {Bernard}, {Bersanelli}, {Bielewicz}, {Bock}, {Bond}, {Borrill}, {Bouchet}, {Boulanger}, {Bucher}, {Burigana}, {Calabrese}, {Cardoso}, {Carron}, {Challinor}, {Chiang}, {Colombo}, {Combet}, {Crill}, {Cuttaia}, {de Bernardis}, {de Zotti}, {Delabrouille}, {Di Valentino}, {Diego}, {Dor{\'e}}, {Douspis}, {Ducout}, {Dupac}, {Efstathiou}, {Elsner}, {En{\ss}lin}, {Eriksen}, {Fantaye}, {Fernandez-Cobos}, {Finelli}, {Forastieri}, {Frailis}, {Fraisse}, {Franceschi}, {Frolov}, {Galeotta}, {Galli}, {Ganga}, {G{\'e}nova-Santos}, {Gerbino}, {Ghosh}, {Gonz{\'a}lez-Nuevo}, {G{\'o}rski}, {Gratton}, {Gruppuso}, {Gudmundsson}, {Hamann}, {Handley}, {Hansen}, {Herranz}, {Hivon}, {Huang}, {Jaffe}, {Jones}, {Karakci}, {Keih{\"a}nen}, {Keskitalo}, {Kiiveri}, {Kim}, {Knox}, {Krachmalnicoff}, {Kunz}, {Kurki-Suonio}, {Lagache}, {Lamarre}, {Lasenby}, {Lattanzi}, {Lawrence}, {Le Jeune}, {Levrier}, {Lewis}, {Liguori}, {Lilje}, {Lindholm}, {L{\'o}pez-Caniego}, {Lubin}, {Ma}, {Mac{\'\i}as-P{\'e}rez}, {Maggio}, {Maino}, {Mandolesi}, {Mangilli}, {Marcos-Caballero}, {Maris}, {Martin}, {Mart{\'\i}nez-Gonz{\'a}lez}, {Matarrese}, {Mauri}, {McEwen}, {Melchiorri}, {Mennella}, {Migliaccio}, {Miville-Desch{\^e}nes}, {Molinari}, {Moneti}, {Montier}, {Morgante}, {Moss}, {Natoli}, {Pagano}, {Paoletti}, {Partridge}, {Patanchon}, {Perrotta}, {Pettorino}, {Piacentini}, {Polastri}, {Polenta}, {Puget}, {Rachen}, {Reinecke}, {Remazeilles}, {Renzi}, {Rocha}, {Rosset}, {Roudier}, {Rubi{\~n}o-Mart{\'\i}n}, {Ruiz-Granados}, {Salvati}, {Sandri}, {Savelainen}, {Scott}, {Sirignano}, {Sunyaev}, {Suur-Uski}, {Tauber}, {Tavagnacco}, {Tenti}, {Toffolatti}, {Tomasi}, {Trombetti}, {Valiviita}, {Van Tent}, {Vielva}, {Villa}, {Vittorio}, {Wandelt}, {Wehus}, {White}, {White}, {Zacchei}, \& {Zonca}}]{collaboration2020planck}
{Planck Collaboration}, {Aghanim}, N., {Akrami}, Y., {et~al.} 2020, \aap, 641, A8, \dodoi{10.1051/0004-6361/201833886}

\bibitem[{{Prochaska} \& {Zheng}(2019)}]{prochaska2019probing}
{Prochaska}, J.~X., \& {Zheng}, Y. 2019, \mnras, 485, 648, \dodoi{10.1093/mnras/stz261}

\bibitem[{{Rafiei-Ravandi} {et~al.}(2021){Rafiei-Ravandi}, {Smith}, {Li}, {Masui}, {Josephy}, {Dobbs}, {Lang}, {Bhardwaj}, {Patel}, {Bandura}, {Berger}, {Boyle}, {Brar}, {Cassanelli}, {Chawla}, {Dong}, {Fonseca}, {Gaensler}, {Giri}, {Good}, {Halpern}, {Kaczmarek}, {Kaspi}, {Leung}, {Lin}, {Mena-Parra}, {Meyers}, {Michilli}, {M{\"u}nchmeyer}, {Ng}, {Petroff}, {Pleunis}, {Rahman}, {Sanghavi}, {Scholz}, {Shin}, {Stairs}, {Tendulkar}, {Vanderlinde}, \& {Zwaniga}}]{rafiei_ravandi2021chime}
{Rafiei-Ravandi}, M., {Smith}, K.~M., {Li}, D., {et~al.} 2021, arXiv e-prints, arXiv:2106.04354.
\newblock \doarXiv{2106.04354}

\bibitem[{{Ravi} {et~al.}(2023){Ravi}, {Catha}, {Chen}, {Connor}, {Cordes}, {Faber}, {Lamb}, {Hallinan}, {Harnach}, {Hellbourg}, {Hobbs}, {Hodge}, {Hodges}, {Law}, {Rasmussen}, {Sharma}, {Sherman}, {Shi}, {Simard}, {Somalwar}, {Squillace}, {Weinreb}, {Woody}, \& {Yadlapalli}}]{ravi2023deep}
{Ravi}, V., {Catha}, M., {Chen}, G., {et~al.} 2023, arXiv e-prints, arXiv:2301.01000, \dodoi{10.48550/arXiv.2301.01000}

\bibitem[{{Reischke} {et~al.}(2024){Reischke}, {Kova{\v{c}}}, {Nicola}, {Hagstotz}, \& {Schneider}}]{reischke2024analytical}
{Reischke}, R., {Kova{\v{c}}}, M., {Nicola}, A., {Hagstotz}, S., \& {Schneider}, A. 2024, arXiv e-prints, arXiv:2411.17682, \dodoi{10.48550/arXiv.2411.17682}

\bibitem[{{Reischke} {et~al.}(2023){Reischke}, {Neumann}, {Bertmann}, {Hagstotz}, \& {Hildebrandt}}]{reischke2023calibrating}
{Reischke}, R., {Neumann}, D., {Bertmann}, K.~A., {Hagstotz}, S., \& {Hildebrandt}, H. 2023, arXiv e-prints, arXiv:2309.09766, \dodoi{10.48550/arXiv.2309.09766}

\bibitem[{{Rudd} {et~al.}(2008){Rudd}, {Zentner}, \& {Kravtsov}}]{rudd2008effects}
{Rudd}, D.~H., {Zentner}, A.~R., \& {Kravtsov}, A.~V. 2008, \apj, 672, 19, \dodoi{10.1086/523836}

\bibitem[{{Seebeck} {et~al.}(2021){Seebeck}, {Ravi}, {Connor}, {Law}, {Simard}, \& {Uzgil}}]{seebeck2021effects}
{Seebeck}, J., {Ravi}, V., {Connor}, L., {et~al.} 2021, arXiv e-prints, arXiv:2112.07639, \dodoi{10.48550/arXiv.2112.07639}

\bibitem[{{Shah} {et~al.}(2024){Shah}, {Shin}, {Leung}, {Fong}, {Eftekhari}, {Amiri}, {Andersen}, {Andrew}, {Bhardwaj}, {Brar}, {Cassanelli}, {Chatterjee}, {Curtin}, {Dobbs}, {Dong}, {Dong}, {Fonseca}, {Gaensler}, {Halpern}, {Hessels}, {Ibik}, {Jain}, {Joseph}, {Kaczmarek}, {Kahinga}, {Kaspi}, {Kharel}, {Landecker}, {Lanman}, {Lazda}, {Main}, {Mas-Ribas}, {Masui}, {Mckinven}, {Mena-Parra}, {Meyers}, {Michilli}, {Nimmo}, {Pandhi}, {Patil}, {Pearlman}, {Pleunis}, {Prochaska}, {Rafiei-Ravandi}, {Sammons}, {Sand}, {Scholz}, {Smith}, \& {Stairs}}]{shah2024repeating}
{Shah}, V., {Shin}, K., {Leung}, C., {et~al.} 2024, arXiv e-prints, arXiv:2410.23374, \dodoi{10.48550/arXiv.2410.23374}

\bibitem[{{Shannon} {et~al.}(2018){Shannon}, {Macquart}, {Bannister}, {Ekers}, {James}, {Os{\l}owski}, {Qiu}, {Sammons}, {Hotan}, {Voronkov}, {Beresford}, {Brothers}, {Brown}, {Bunton}, {Chippendale}, {Haskins}, {Leach}, {Marquarding}, {McConnell}, {Pilawa}, {Sadler}, {Troup}, {Tuthill}, {Whiting}, {Allison}, {Anderson}, {Bell}, {Collier}, {G{\"u}rkan}, {Heald}, \& {Riseley}}]{shannon2018dispersion}
{Shannon}, R.~M., {Macquart}, J.~P., {Bannister}, K.~W., {et~al.} 2018, \nat, 562, 386, \dodoi{10.1038/s41586-018-0588-y}

\bibitem[{{Shannon} {et~al.}(2024){Shannon}, {Bannister}, {Bera}, {Bhandari}, {Day}, {Deller}, {Dial}, {Dobie}, {Ekers}, {Fong}, {Glowacki}, {Gordon}, {Gourdji}, {Jaini}, {James}, {Kumar}, {Mahony}, {Marnoch}, {Muller}, {Prochaska}, {Qiu}, {Ryder}, {Sadler}, {Scott}, {Tejos}, {Uttarkar}, \& {Wang}}]{shannon2024commensal}
{Shannon}, R.~M., {Bannister}, K.~W., {Bera}, A., {et~al.} 2024, arXiv e-prints, arXiv:2408.02083, \dodoi{10.48550/arXiv.2408.02083}

\bibitem[{{Sharma} {et~al.}(2025){Sharma}, {Krause}, {Ravi}, {Reischke}, {Pranjal R.}, \& {Connor}}]{sharma2025hydrodynamical}
{Sharma}, K., {Krause}, E., {Ravi}, V., {et~al.} 2025, arXiv e-prints, arXiv:2504.18745, \dodoi{10.48550/arXiv.2504.18745}

\bibitem[{{Sharma} {et~al.}(2023){Sharma}, {Somalwar}, {Law}, {Ravi}, {Catha}, {Chen}, {Connor}, {Faber}, {Hallinan}, {Harnach}, {Hellbourg}, {Hobbs}, {Hodge}, {Hodges}, {Lamb}, {Rasmussen}, {Sherman}, {Shi}, {Simard}, {Squillace}, {Weinreb}, {Woody}, \& {Yadlapalli}}]{sharma2023deep}
{Sharma}, K., {Somalwar}, J., {Law}, C., {et~al.} 2023, arXiv e-prints, arXiv:2302.14782, \dodoi{10.48550/arXiv.2302.14782}

\bibitem[{{Sharma} {et~al.}(2024){Sharma}, {Ravi}, {Connor}, {Law}, {Ocker}, {Sherman}, {Kosogorov}, {Faber}, {Hallinan}, {Harnach}, {Hellbourg}, {Hobbs}, {Hodge}, {Hodges}, {Lamb}, {Rasmussen}, {Somalwar}, {Weinreb}, {Woody}, {Leja}, {Anand}, {Das}, {Qin}, {Rose}, {Dong}, {Miller}, \& {Yao}}]{sharma2024preferential}
{Sharma}, K., {Ravi}, V., {Connor}, L., {et~al.} 2024, \nat, 635, 61, \dodoi{10.1038/s41586-024-08074-9}

\bibitem[{{Shin} {et~al.}(2023){Shin}, {Masui}, {Bhardwaj}, {Cassanelli}, {Chawla}, {Dobbs}, {Dong}, {Fonseca}, {Gaensler}, {Herrera-Mart{\'\i}n}, {Kaczmarek}, {Kaspi}, {Leung}, {Merryfield}, {Michilli}, {M{\"u}nchmeyer}, {Pearlman}, {Rafiei-Ravandi}, {Smith}, {Stairs}, \& {Tendulkar}}]{shin2023inferring}
{Shin}, K., {Masui}, K.~W., {Bhardwaj}, M., {et~al.} 2023, \apj, 944, 105, \dodoi{10.3847/1538-4357/acaf06}

\bibitem[{{Shirasaki} {et~al.}(2017){Shirasaki}, {Kashiyama}, \& {Yoshida}}]{shirasaki2017large}
{Shirasaki}, M., {Kashiyama}, K., \& {Yoshida}, N. 2017, \prd, 95, 083012, \dodoi{10.1103/PhysRevD.95.083012}

\bibitem[{{Shirasaki} {et~al.}(2022){Shirasaki}, {Takahashi}, {Osato}, \& {Ioka}}]{shirasaki2022probing}
{Shirasaki}, M., {Takahashi}, R., {Osato}, K., \& {Ioka}, K. 2022, \mnras, 512, 1730, \dodoi{10.1093/mnras/stac490}

\bibitem[{{Simha} {et~al.}(2020){Simha}, {Burchett}, {Prochaska}, {Chittidi}, {Elek}, {Tejos}, {Jorgenson}, {Bannister}, {Bhandari}, {Day}, {Deller}, {Forbes}, {Macquart}, {Ryder}, \& {Shannon}}]{simha2020disentangling}
{Simha}, S., {Burchett}, J.~N., {Prochaska}, J.~X., {et~al.} 2020, \apj, 901, 134, \dodoi{10.3847/1538-4357/abafc3}

\bibitem[{{Simha} {et~al.}(2021){Simha}, {Tejos}, {Prochaska}, {Lee}, {Ryder}, {Cantalupo}, {Bannister}, {Bhandari}, \& {Shannon}}]{simha2021estimating}
{Simha}, S., {Tejos}, N., {Prochaska}, J.~X., {et~al.} 2021, \apj, 921, 134, \dodoi{10.3847/1538-4357/ac2000}

\bibitem[{{Spergel} {et~al.}(2015){Spergel}, {Gehrels}, {Baltay}, {Bennett}, {Breckinridge}, {Donahue}, {Dressler}, {Gaudi}, {Greene}, {Guyon}, {Hirata}, {Kalirai}, {Kasdin}, {Macintosh}, {Moos}, {Perlmutter}, {Postman}, {Rauscher}, {Rhodes}, {Wang}, {Weinberg}, {Benford}, {Hudson}, {Jeong}, {Mellier}, {Traub}, {Yamada}, {Capak}, {Colbert}, {Masters}, {Penny}, {Savransky}, {Stern}, {Zimmerman}, {Barry}, {Bartusek}, {Carpenter}, {Cheng}, {Content}, {Dekens}, {Demers}, {Grady}, {Jackson}, {Kuan}, {Kruk}, {Melton}, {Nemati}, {Parvin}, {Poberezhskiy}, {Peddie}, {Ruffa}, {Wallace}, {Whipple}, {Wollack}, \& {Zhao}}]{spergel2015wide}
{Spergel}, D., {Gehrels}, N., {Baltay}, C., {et~al.} 2015, arXiv e-prints, arXiv:1503.03757, \dodoi{10.48550/arXiv.1503.03757}

\bibitem[{{Spitler} {et~al.}(2016){Spitler}, {Scholz}, {Hessels}, {Bogdanov}, {Brazier}, {Camilo}, {Chatterjee}, {Cordes}, {Crawford}, {Deneva}, {Ferdman}, {Freire}, {Kaspi}, {Lazarus}, {Lynch}, {Madsen}, {McLaughlin}, {Patel}, {Ransom}, {Seymour}, {Stairs}, {Stappers}, {van Leeuwen}, \& {Zhu}}]{spitler2016repeating}
{Spitler}, L.~G., {Scholz}, P., {Hessels}, J.~W.~T., {et~al.} 2016, \nat, 531, 202, \dodoi{10.1038/nature17168}

\bibitem[{{Tempel} {et~al.}(2018){Tempel}, {Kruuse}, {Kipper}, {Tuvikene}, {Sorce}, \& {Stoica}}]{tempel2018bayesian}
{Tempel}, E., {Kruuse}, M., {Kipper}, R., {et~al.} 2018, \aap, 618, A81, \dodoi{10.1051/0004-6361/201833217}

\bibitem[{{Tendulkar} {et~al.}(2017){Tendulkar}, {Bassa}, {Cordes}, {Bower}, {Law}, {Chatterjee}, {Adams}, {Bogdanov}, {Burke-Spolaor}, {Butler}, {Demorest}, {Hessels}, {Kaspi}, {Lazio}, {Maddox}, {Marcote}, {McLaughlin}, {Paragi}, {Ransom}, {Scholz}, {Seymour}, {Spitler}, {van Langevelde}, \& {Wharton}}]{tendulkar2017host}
{Tendulkar}, S.~P., {Bassa}, C.~G., {Cordes}, J.~M., {et~al.} 2017, \apjl, 834, L7, \dodoi{10.3847/2041-8213/834/2/L7}

\bibitem[{{Terasawa} {et~al.}(2025){Terasawa}, {Li}, {Takada}, {Nishimichi}, {Tanaka}, {Sugiyama}, {Kurita}, {Zhang}, {Shirasaki}, {Takahashi}, {Miyatake}, {More}, \& {Nishizawa}}]{terasawa2025exploring}
{Terasawa}, R., {Li}, X., {Takada}, M., {et~al.} 2025, \prd, 111, 063509, \dodoi{10.1103/PhysRevD.111.063509}

\bibitem[{{Tr{\"o}ster} {et~al.}(2022){Tr{\"o}ster}, {Mead}, {Heymans}, {Yan}, {Alonso}, {Asgari}, {Bilicki}, {Dvornik}, {Hildebrandt}, {Joachimi}, {Kannawadi}, {Kuijken}, {Schneider}, {Shan}, {van Waerbeke}, \& {Wright}}]{tr_oster2022joint}
{Tr{\"o}ster}, T., {Mead}, A.~J., {Heymans}, C., {et~al.} 2022, \aap, 660, A27, \dodoi{10.1051/0004-6361/202142197}

\bibitem[{{Tully}(2015)}]{tully2015galaxy}
{Tully}, R.~B. 2015, \aj, 149, 171, \dodoi{10.1088/0004-6256/149/5/171}

\bibitem[{{Turk} {et~al.}(2011){Turk}, {Smith}, {Oishi}, {Skory}, {Skillman}, {Abel}, \& {Norman}}]{turk2011yt}
{Turk}, M.~J., {Smith}, B.~D., {Oishi}, J.~S., {et~al.} 2011, \apjs, 192, 9, \dodoi{10.1088/0067-0049/192/1/9}

\bibitem[{{Vanderlinde} {et~al.}(2019){Vanderlinde}, {Liu}, {Gaensler}, {Bond}, {Hinshaw}, {Ng}, {Chiang}, {Stairs}, {Brown}, {Sievers}, {Mena}, {Smith}, {Bandura}, {Masui}, {Spekkens}, {Belostotski}, {Dobbs}, {Turok}, {Boyle}, {Rupen}, {Landecker}, {Pen}, \& {Kaspi}}]{vanderlinde2019canadian}
{Vanderlinde}, K., {Liu}, A., {Gaensler}, B., {et~al.} 2019, in Canadian Long Range Plan for Astronomy and Astrophysics White Papers, Vol. 2020, 28, \dodoi{10.5281/zenodo.3765414}

\bibitem[{{Villaescusa-Navarro} {et~al.}(2021){Villaescusa-Navarro}, {Angl{\'e}s-Alc{\'a}zar}, {Genel}, {Spergel}, {Somerville}, {Dave}, {Pillepich}, {Hernquist}, {Nelson}, {Torrey}, {Narayanan}, {Li}, {Philcox}, {La Torre}, {Maria Delgado}, {Ho}, {Hassan}, {Burkhart}, {Wadekar}, {Battaglia}, {Contardo}, \& {Bryan}}]{camels_presentation}
{Villaescusa-Navarro}, F., {Angl{\'e}s-Alc{\'a}zar}, D., {Genel}, S., {et~al.} 2021, \apj, 915, 71, \dodoi{10.3847/1538-4357/abf7ba}

\bibitem[{{Villaescusa-Navarro} {et~al.}(2023){Villaescusa-Navarro}, {Genel}, {Angl{\'e}s-Alc{\'a}zar}, {Perez}, {Villanueva-Domingo}, {Wadekar}, {Shao}, {Mohammad}, {Hassan}, {Moser}, {Lau}, {Machado Poletti Valle}, {Nicola}, {Thiele}, {Jo}, {Philcox}, {Oppenheimer}, {Tillman}, {Hahn}, {Kaushal}, {Pisani}, {Gebhardt}, {Delgado}, {Caliendo}, {Kreisch}, {Wong}, {Coulton}, {Eickenberg}, {Parimbelli}, {Ni}, {Steinwandel}, {La Torre}, {Dave}, {Battaglia}, {Nagai}, {Spergel}, {Hernquist}, {Burkhart}, {Narayanan}, {Wandelt}, {Somerville}, {Bryan}, {Viel}, {Li}, {Irsic}, {Kraljic}, {Marinacci}, \& {Vogelsberger}}]{camels_data_release}
{Villaescusa-Navarro}, F., {Genel}, S., {Angl{\'e}s-Alc{\'a}zar}, D., {et~al.} 2023, \apjs, 265, 54, \dodoi{10.3847/1538-4365/acbf47}

\bibitem[{{Virtanen} {et~al.}(2020){Virtanen}, {Gommers}, {Oliphant}, {Haberland}, {Reddy}, {Cournapeau}, {Burovski}, {Peterson}, {Weckesser}, {Bright}, {van der Walt}, {Brett}, {Wilson}, {Jarrod Millman}, {Mayorov}, {Nelson}, {Jones}, {Kern}, {Larson}, {Carey}, {Polat}, {Feng}, {Moore}, {Vand erPlas}, {Laxalde}, {Perktold}, {Cimrman}, {Henriksen}, {Quintero}, {Harris}, {Archibald}, {Ribeiro}, {Pedregosa}, {van Mulbregt}, \& {Contributors}}]{virtanen2020scipy}
{Virtanen}, P., {Gommers}, R., {Oliphant}, T.~E., {et~al.} 2020, Nature Methods, 17, 261, \dodoi{https://doi.org/10.1038/s41592-019-0686-2}

\bibitem[{{Walker} {et~al.}(2024){Walker}, {Spitler}, {Ma}, {Cheng}, {Artale}, \& {Hummels}}]{walker2024dispersion}
{Walker}, C. R.~H., {Spitler}, L.~G., {Ma}, Y.-Z., {et~al.} 2024, \aap, 683, A71, \dodoi{10.1051/0004-6361/202347139}

\bibitem[{{Wang} {et~al.}(2025{\natexlab{a}}){Wang}, {Masui}, {Andrew}, {Fonseca}, {Gaensler}, {Joseph}, {Kaspi}, {Kharel}, {Lanman}, {Leung}, {Mas-Ribas}, {Mena-Parra}, {Nimmo}, {Pearlman}, {Pen}, {Prochaska}, {Raikman}, {Shin}, {Siegel}, {Smith}, \& {Stairs}}]{wang2025measurement}
{Wang}, H., {Masui}, K., {Andrew}, S., {et~al.} 2025{\natexlab{a}}, arXiv e-prints, arXiv:2506.08932, \dodoi{10.48550/arXiv.2506.08932}

\bibitem[{{Wang} {et~al.}(2025{\natexlab{b}}){Wang}, {Bannister}, {Gupta}, {Deng}, {Pilawa}, {Tuthill}, {Bunton}, {Flynn}, {Glowacki}, {Jaini}, {Lee}, {Lenc}, {Lucero}, {Paek}, {Radhakrishnan}, {Thyagarajan}, {Uttarkar}, {Wang}, {Bhat}, {James}, {Moss}, {Murphy}, {Reynolds}, {Shannon}, {Spitler}, {Tzioumis}, {Caleb}, {Deller}, {Gordon}, {Marnoch}, {Ryder}, {Simha}, {Anderson}, {Ball}, {Brodrick}, {Cooray}, {Gupta}, {Hayman}, {Ng}, {Pearce}, {Phillips}, {Voronkov}, \& {Westmeier}}]{wang2025craft}
{Wang}, Z., {Bannister}, K.~W., {Gupta}, V., {et~al.} 2025{\natexlab{b}}, \pasa, 42, e005, \dodoi{10.1017/pasa.2024.107}

\bibitem[{{Wen} {et~al.}(2018){Wen}, {Han}, \& {Yang}}]{wen2018catalogue}
{Wen}, Z.~L., {Han}, J.~L., \& {Yang}, F. 2018, \mnras, 475, 343, \dodoi{10.1093/mnras/stx3189}

\bibitem[{{White}(2004)}]{white2004baryons}
{White}, M. 2004, Astroparticle Physics, 22, 211, \dodoi{10.1016/j.astropartphys.2004.06.001}

\bibitem[{{Wright} {et~al.}(2024){Wright}, {Somerville}, {Lagos}, {Schaller}, {Dav{\'e}}, {Angl{\'e}s-Alc{\'a}zar}, \& {Genel}}]{wright2024baryon}
{Wright}, R.~J., {Somerville}, R.~S., {Lagos}, C. d.~P., {et~al.} 2024, \mnras, 532, 3417, \dodoi{10.1093/mnras/stae1688}

\bibitem[{{Xu} {et~al.}(2022){Xu}, {Ramos-Ceja}, {Pacaud}, {Reiprich}, \& {Erben}}]{xu2022catalog}
{Xu}, W., {Ramos-Ceja}, M.~E., {Pacaud}, F., {Reiprich}, T.~H., \& {Erben}, T. 2022, \aap, 658, A59, \dodoi{10.1051/0004-6361/202140908}

\bibitem[{{Yamasaki} \& {Totani}(2020)}]{yamasaki2020galactic}
{Yamasaki}, S., \& {Totani}, T. 2020, \apj, 888, 105, \dodoi{10.3847/1538-4357/ab58c4}

\bibitem[{{Zhan} \& {Knox}(2004)}]{zhan2004effect}
{Zhan}, H., \& {Knox}, L. 2004, \apjl, 616, L75, \dodoi{10.1086/426712}

\end{thebibliography}



\end{document}